\documentclass[aps,prx,twocolumn,%superscriptaddress,
groupedaddress,longbibliography]{revtex4-2}

\usepackage{dcolumn}
\usepackage{url}
\expandafter\let\csname equation*\endcsname\relax
\expandafter\let\csname endequation*\endcsname\relax
\usepackage{amsmath,amssymb}
\usepackage{amsthm}
\usepackage{graphicx,bm,color,hyperref}
\usepackage{mathrsfs}
\usepackage{dsfont}
\usepackage{mathtools}
\usepackage{bbm,bm}
\usepackage{mathdots}
\usepackage{xcolor}
\usepackage{tabularx}
\usepackage{booktabs}
\usepackage{hyperref}
\usepackage{amssymb}
\usepackage{amsfonts}
\usepackage{changes}
\usepackage{subfigure}
\usepackage{xcolor}
\usepackage{float}
\usepackage{cleveref}
\usepackage{enumitem}

\usepackage{mhchem}

\usepackage{tikz}

\usepackage{changes}

\usepackage{cleveref}
\crefname{theorem}{Theorem}{Theorems}
\crefname{proposition}{Proposition}{Propositions}
\crefname{definition}{Definition}{Definitions}
\crefname{lemma}{Lemma}{Lemmas}
\crefname{figure}{Figure}{Figures}
\crefname{corollary}{Corollary}{Corollary}
\crefname{conjecture}{Conjecture}{Conjectures}
\crefname{section}{Section}{Sections}
\crefname{appendix}{Appendix}{Appendixes}
\crefname{observation}{Observation}{Observation}
\crefname{remark}{Remark}{Remark}
\crefname{example}{Example}{Examples}
\crefname{equation}{Eq.}{Eqs.}
\crefname{table}{Table}{Tables}
\crefname{theorem}{Theorem}{Theorems}

\newcommand\dloc{q} %local dimension
\DeclareMathOperator{\hiH}{\mathcal{H}} 

\DeclareMathOperator{\e}{\mathrm{e}}
\DeclareMathOperator{\iu}{\mathrm{i}}
\DeclareMathOperator{\Tr}{\mathrm{Tr}}

\newcommand{\ket}[1]{\vert #1 \rangle}
\newcommand{\bra}[1]{\langle #1 \vert}
\newcommand{\ketbra}[2]{\vert #1 \rangle\langle #2\vert}

\newcommand{\floor}[1]{\left\lfloor #1 \right\rfloor}
\newcommand\AME{\mathrm{AME}}

\newcommand{\1}{\ensuremath{\mathbbm{1}}}
\renewcommand{\emph}{\textit}

\newcommand{\qd}{\text{e}} %
\newcommand{\p}{\text{p}}
\newcommand{\step}[1]{_\text{step$#1$}}

\begin{document}

\title{Deterministic generation of qudit photonic graph states from quantum emitters}

 \author{Zahra Raissi}
 \affiliation{Department of Computer Science and Institute for Photonic Quantum Systems (PhoQS), Paderborn University, Germany}
 \affiliation{Department of Physics, Virginia Tech, Blacksburg, VA 24061, USA}
\affiliation{Virginia Tech Center for Quantum Information Science and Engineering, Blacksburg, VA 24061, USA}

\author{Edwin Barnes, and Sophia E. Economou}
 \affiliation{Department of Physics, Virginia Tech, Blacksburg, VA 24061, USA}
\affiliation{Virginia Tech Center for Quantum Information Science and Engineering, Blacksburg, VA 24061, USA}

\begin{abstract} 
We propose and analyze deterministic protocols to generate qudit photonic graph states from quantum emitters. {We show that our approach can be applied to generate any qudit graph state, and we exemplify it} by constructing protocols to generate {one- and two-dimensional qudit cluster states}, absolutely maximally entangled states, and logical states of quantum error correcting codes. Some of these protocols make use of time-delayed feedback, while others do not. {The only additional resource requirement compared to the qubit case is the ability to control multi-level emitters.} These results significantly broaden the range of multi-photon entangled states that can be produced deterministically from quantum emitters.
\end{abstract} 

\maketitle

\section{Introduction}

Entanglement is a uniquely quantum property that plays an important role in almost all aspects of quantum information science, including quantum computing \cite{Horodecki-general-paper},
quantum error correction \cite{Scott,Calderbank}, quantum sensing \cite{DegenRMP2017}, and quantum networks \cite{Cirac-QNetwork,Kimble-QInternet,Hillery-Secret-Sharing}. Many of these applications require generating large multi-photon entangled resource states upfront, especially in the context of measurement- or fusion-based quantum computing \cite{Raussendorf-measurement-based,Bartolucci_arxiv2021,Omkar-multiphoton} and quantum communication \cite{Gisin2007,Azuma2015,Zwerger2016,
Muralidharan2016}.

However, creating entangled states of many photons is challenging because photons do not interact directly.
Standard ways of circumventing this issue make use of either nonlinear media \cite{Guerreiro2014} or quantum interference and measurement \cite{Knill-multiphoton, Kok-photonsystem}; the former approach is made challenging by low coupling efficiencies, while the latter is intrinsically probabilistic. Approaches that rely on interfering photons are usually based on linear optics and post-selection \cite{Knill-multiphoton, Kok-photonsystem}, and consequently the success probability decreases exponentially with the number of photons \cite{Knill-multiphoton, bodiya2006scalable}.
Despite a number of conceptual and technological advances, the probabilistic nature of this approach continues to severely limit the size of multi-photon entangled states constructed in this way \cite{Gao2010,LiACS2020}.

An alternative approach is to use coupled, controllable quantum emitters with suitable level structures to deterministically generate multi-photon entanglement \cite{SchonPRL2005,Lindner2009,Economou2010}.
There now exist several explicit protocols for creating entangled states of many photonic qubits, either by using entangling gates between emitters and transferring entanglement to the photons in the photon emission stage \cite{Economou2010,ButerakosPRX2017,
Russophotonic-graph, GSegovia2019,Hilaire2021resource,
Liu_arxiv2022,Li2022}, or by sending the photons to interact again with emitters to potentially create entanglement beyond what is generated from the emission process \cite{wiseman_milburn_2009,
Pichler2017,Zhan_PRL2020,Yu-timefeedback}. Proof-of-principle experimental demonstrations of such deterministic protocols have been performed in both the optical and microwave domains \cite{Schwartz2016,Besse2020,Rempe-Nature2022,Coste_2022}.

To date, the vast majority of theoretical and experimental efforts towards the deterministic generation of multi-photon entangled states have focused on photonic qubits. These are based on using either polarization, spatial path, or time bin as the logical encoding. However, photons can naturally encode not only qubits but also multi-dimensional qudit states, for example by using more than two spatial paths or time bins. This can allow for novel approaches to quantum computing, communication, sensing, and error correction in which quantum information is stored in a more compact way \cite{Kues-qudits, Erhard-qudits, Wang-qudit}. Such states can in particular provide benefits in quantum networks and repeaters~\cite{Muralidharan_PRA2018,Zheng_2022}.  Although there have been recent experimental demonstrations of entangled qudit state creation, these have been limited thus far to two photons \cite{Kues-qudits,Reimer2019}. An outstanding question is whether multi-photon entangled qudit states can be generated deterministically from a small number of quantum emitters.

In this paper, we propose and evaluate deterministic methods to generate multi-photon qudit graph states from multi-level quantum emitters. We present several different explicit protocols that can produce various states either using a single emitter together with time-delayed feedback, or using multiple coupled quantum emitters.
{We first show that any qudit graph state can be produced from multi-level quantum emitters with an appropriate level structure using a small set of gates on the emitters. We then focus on constructing one- and two-dimensional cluster states, as well as highly entangled multipartite states called absolutely maximally entangled (AME) states.}
These states are defined by the property that they maximize the entanglement entropy for any bipartition \cite{Helwig2012,Helwig2013}, and they have applications in quantum error correcting codes (QECCs) and secret sharing \cite{Scott,Helwig2012,Helwig2013,Zahra-minimalsupport, Zahra-QOA, Zahra-CL+Q, Zahra-MShortening, Zahra-hierarchy}.
In addition, we present protocols for constructing logical states of QECCs whose code spaces are spanned by AME states of qutrits {Our approach to qudit photonic graph state generation incurs only a small resource overhead compared to the qubit case, primarily in the requirement of multi-level control of quantum emitters, which has been achieved experimentally in atomic and defect-based systems~\cite{DOHERTY20131,Wolfowicz2021,Ringbauer-ION-2021,Aksenov-ION-2022}.}

The paper is organized as follows. We begin with a brief review of qudit graph states and basic qudit operations (Sec.~\ref{sec:Qudits-and-graph}). 
In Sec.~\ref{sec:Qutrit-clusterstates}, we describe how to produce photonic qudits from quantum emitters.
{In Sec.~\ref{sec:graph-state-generation}, we illustrate our basic approach to multi-qudit-state generation and show that any qudit photonic graph state can be generated from quantum emitters using a small set of operations. We also present examples of generating one and two-dimensional qudit graph states.}
In Sec.~\ref{sec:AME-states}, we present protocols for generating various AME states.
In Sec.~\ref{sec:QECCs}, we show how to generate an explicit example of a QECC whose codewords are all AME states of qutrits. Sec.~\ref{sec:implementations} discusses possible physical implementations. We conclude in Sec.~\ref{sec:conclusion}. 

\section{Background: qudits and graph states}\label{sec:Qudits-and-graph}

In this work, we focus on the generation of an important class of entangled states called graph states \cite{Hein-2006}.
Graph states are pure quantum states that are defined based on a graph $G=(V, \Gamma)$, which is composed of a set $V$ of $n$ vertices (each qudit is represented by a vertex), and a set of edges specified by the adjacency matrix $\Gamma$.
$\Gamma$ is an $n \times n$ symmetric matrix such that $\Gamma_{i,j}=0$ if vertices $i$ and $j$ are not connected and $\Gamma_{i,j} >0$ otherwise \cite{Nest,Hein,Hein-2006,Bahramgiri}. These states have many applications in measurement-based quantum computing \cite{Raussendorf-1wayQC,Bartolucci_arxiv2021}, quantum networks \cite{Azuma2015,Zwerger2016,Muralidharan2016}, and QECCs \cite{Gottesman-thesis,Gottesman}.
 
There are differences between qubit and qudit graph states. 
For the qubit case, the graph state is defined by initializing all qubits in the $+1$ eigenstate of the $X$ Pauli matrix, i.e., the state $\ket{+}= \ket 0+\ket 1$ (here and in the following we will
not always explicitly normalize states for the sake of a more compact notation), and then applying controlled-$Z$ ($CZ$) gates on all pairs of qubits connected by an edge. In the case of qubits, the adjacency matrix $\Gamma$ contains two different elements: $\Gamma_{i,j}=0$ whenever two vertices $i$ and $j$ are not connected, and $\Gamma_{i,j}=1$ otherwise. Most of the existing protocols for creating multi-photon entangled states have focused on the generation of multi-qubit graph states \cite{Lindner2009,Economou2010,
ButerakosPRX2017,Russophotonic-graph, GSegovia2019,Hilaire2021resource,
Liu_arxiv2022,Li2022,Pichler2017,
Zhan_PRL2020,Yu-timefeedback}.

To define qudit graph states, we first introduce generalized Pauli operators acting on qudits with $\dloc$ levels \cite{Bahramgiri}.
The Pauli operators $X$ and $Z$ act on the eigenstates of $Z$ as follows: 
\begin{align} 
  X\ket{i}&=\ket{i+1 \mod \dloc}, \label{eq:defX}\\
  Z\ket{i}&=\omega^i\,\ket{i}\, . \label{eq:defZ}
\end{align}
These operators are unitary and traceless, and they satisfy the condition $ZX=\omega XZ$, where $\omega=\e^{\iu 2 \pi/\dloc}$ is a $\dloc$-th root of unity.
Each of the Pauli matrices has $\dloc$ eigenvalues and eigenvectors. 
The powers of $X$ and $Z$ 
applied to basis states give
\begin{align}  
  X^\alpha \ket{i}&=\ket{i+\alpha \mod \dloc}, \label{eq:defXqudit}\\
  Z^\beta\ket{i}&=\omega^{i\beta}\,\ket{i}\, ,  \label{eq:defZqudit}
\end{align}
and $X^\dloc = Z^\dloc = \1$.
Similar to the case of qubits, we can define controlled-$Z^\beta$ ($CZ^\beta$) gates as 
\begin{equation}
  CZ^\beta \ket{i,j} =\omega^{\beta \, ij} \ket{i,j}, \qquad \forall \beta \in \{1,\dots ,\dloc-1 \} \label{eq:CZqudit}\ .
\end{equation}
The special case of $\dloc =2$ in the above corresponds to qubits. 

One other operator that is essential to studying qudit graph states is the  Hadamard gate. 
The Hadamard gate $H$ is a matrix that maps the $Z$-eigenbasis into the $X$-eigenbasis.
For qubits, we have $H \ket{0} = \ket {+}$, and $H \ket{1} = \ket{-}$, where $\ket{-}$ is the $-1$ eigenvector of the $X$ Pauli operator.
There is also a useful identity involving the Hadamard and the Pauli matrices $X$ and $Z$: $HXH=Z$.

Similarly to the qubit case, we can define the qudit Hadamard operator $H$ such that $H \ket{i} = \ket{X_i}$, where $\ket{X_i}$ is an eigenstate of $X$. In general, we can write
 \begin{equation}\label{eq:H-on-qudits}
   H \ket{i} = \sum_{j=0}^{\dloc-1} \omega^{ij} \ket{j} = \ket{X_i}\qquad \forall i \in \{0,1,...,\dloc-1\}\ .
 \end{equation} 
The relationship between the Hadamard and the Pauli operators $X^\alpha$ and $Z^\alpha$ in the case of qudits takes the form 
 \begin{equation}
   HX^\alpha H^\dagger = Z^\alpha, \label{eq:XZH-qudits}
 \end{equation}
where $H^\dag$ is the inverse of $H$:
 \begin{equation}\label{eq:Hdag-on-qudits}
   H^\dag \ket{i} = \sum_{j=0}^{\dloc-1} \omega^{(\dloc-1)ij} \ket{j} = \ket{X_{q-i\!\!\!\!\mod \dloc}}, \quad \forall i=\{0,...,\dloc-1\}
 \end{equation} 
Also note that $H^\dag$ acts on the $X$-eigenbasis as $H^\dag \ket{X_i} = \ket{i}$. 

To understand how we can use these ingredients to define qudit graph states, we present an explicit example involving qutrits.
A qutrit is realized by a 3-level quantum system ($\dloc =3$), where we denote the $Z$-eigenstates by $\ket 0$, $\ket 1$, and $\ket 2$.
The $X$-eigenstates then are
\begin{align} 
  \ket{X_0}&=\ket{0}+\ket{1}+\ket{2}, \label{eq:X0}\\
  \ket{X_1}&=\ket{0}+\omega\ket{1}+\omega^2\ket{2}, \label{eq:X1}\\
 \ket{X_2}&=\ket{0}+\omega^2\ket{1}+\omega\ket{2}\ , \label{eq:X2}
\end{align}
where $\omega=e^{\iu 2\pi/3}$.
The $X$ operator couples levels as follows: $X \ket{i}=\ket{i+1\mod 3}$ and $X^2 \ket{i}=\ket{i+2\mod 3} $.
The $Z$ and $Z^2$ operators add different phase factors to each basis state, i.e., $Z \ket{i} = \omega^{i}\ket{i}$ and $Z^2 \ket{i} = \omega^{2i} \ket{i}$, while $X^3 = \1$ and $Z^3 = \1$.

In order to obtain qutrit graph states, we first initialize all qutrits in state $\ket{X_0}$.
For each edge, we can consider two possible gates on the corresponding qutrits: $CZ$ and $CZ^2$, which are defined such that  
\begin{align} 
  CZ \ket{i,j}&=\omega^{ij} \ket{i,j}, \label{eq:CZqutrit}\\
  CZ^2 \ket{i,j} &=\omega^{2\, ij} \ket{i,j}, \label{eq:CZ2qutrit}
\end{align}
where $i , j \in \{ 0,1,2 \}$.
Therefore, the adjacency matrix $\Gamma$ contains three different elements, $\Gamma_{i,j}=0$ when the two vertices $i$ and $j$ are not connected, $\Gamma_{i,j}=1$ when they are connected via $CZ$, and $\Gamma_{i,j}=2$ when they are connected via $CZ^2$.

{A general graph state of $n$ qudits can be expressed as
\begin{align}\label{eq:general_graph_state}
&\left(\prod_{i<j}CZ_{i,j}^{\Gamma_{i,j}}\right)\ket{X_0}^{\otimes n}\nonumber\\ &=\left(\prod_{i<j}CZ_{i,j}^{\Gamma_{i,j}}\right)\sum_{k_1,k_2,...,k_n=0}^{q-1}\ket{k_1k_2...k_n}\nonumber\\ &=\sum_{k_1,k_2,...,k_n=0}^{q-1}\left(\prod_{i<j}\omega^{k_ik_j\Gamma_{i,j}}\right)\ket{k_1k_2...k_n}.
\end{align}
}

\section{Emitting photonic qudits }
\label{sec:Qutrit-clusterstates}

In addition to qudit operations, another key ingredient we need to generate qudit photonic graph states is a ``pumping" operation that produces photonic qudits from quantum emitters. In this section, we explain how such operations can be realized in systems containing multi-level emitters with the appropriate level structure and selection rules. Later on in Sec.~\ref{sec:implementations}, we give explicit examples of such systems, which include color centers in solids, trapped ions, and neutral atoms. In the present section, we also illustrate how photon pumping, together with qudit operations on emitters and photons, can be used to deterministically create entanglement between photonic qudits.

As a warm-up example illustrating how this works, {in Sec.~\ref{sec:graph-state-generation} below} we present protocols for generating one-dimensional (1D) qudit graph states using a single quantum emitter with an appropriate level structure. This example is closely related to Ref.~\cite{Lindner-Rudolph-2009}, which showed how to produce a qubit 1D graph state from an emitter comprised of a single electron spin in an optically active quantum dot. That work introduced a protocol in which each photon emission is preceded by a Hadamard gate. Repeating the basic sequence of Hadamard gate followed by optical pumping/photon emission $n$ times results in a $n$-qubit linear graph state. Because each photon emission can be viewed as a $CNOT$ gate acting on the emitter and emitted photon, this operation together with the Hadamard gates effectively generates the requisite $CZ$ gate between neighboring photonic qubits.

Although the original proposal of Ref.~\cite{Lindner-Rudolph-2009} focused on using photon polarization as the qubit degree of freedom, it is important to note that one can also generate a linear graph state based on time-bin qubits by using circularly (rather than linearly) polarized light to pump the emitter and by including a few additional operations during each cycle of the protocol \cite{Lee_2019}. In this protocol, only one of the emitter ground states, say $\ket{0}_\qd$, can be optically excited by the pump, so that an initial superposition state $\ket{0}_\qd+\ket{1}_\qd$ becomes $\ket{0}_\qd\ket{0}_\p+\ket{1}_\qd\ket{\text{vac}}$ after the first photon emission, where $\ket{0}_\p$ corresponds to a photon in the first time bin, while $\ket{\text{vac}}$ represents the vacuum.
If we then apply an $X$ gate on the emitter and perform a second optical pumping operation, we obtain $\ket{1}_\qd\ket{0}_\p+\ket{0}_\qd\ket{1}_\p$, where $\ket{1}_\p$ corresponds to a photon in the second time bin.

\begin{figure}
\includegraphics[width=\columnwidth]{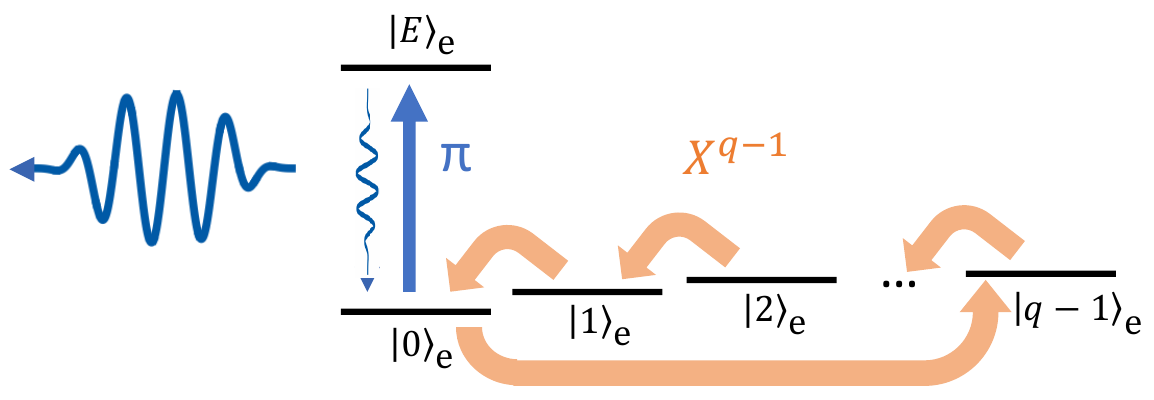}
\caption{Photon-pumping operation ${\cal P}_\mathrm{pump}$. Time-bin photonic qudits with $q$ levels are generated from a quantum emitter with $q$ ground states ($\ket{0}_\mathrm{e},\ldots,\ket{q-1}_\mathrm{e}$), one of which ($\ket{0}_\mathrm{e}$) couples to an optically excited state ($\ket{E}_\mathrm{e}$). A time-bin photonic qudit is created by applying optical $\pi$ pulses and $X^{q-1}$ gates in alternating fashion. Each  $X^{q-1}$ gate rotates the $q$ ground states cyclically as shown. By applying the $\pi$-pulse/$X^{q-1}$ gate sequence $q$ times, one obtains a photonic time-bin qudit with $q$ bins that is entangled with the emitter.}\label{fig:photon-pumping}
\end{figure}

The above protocol for time-bin qubits naturally extends to time-bin qudits with local dimension $q$, provided we have at our disposal a quantum emitter with $q$ energy levels comprising the ground state manifold, one of which is optically coupled to an excited state. 
As in the qubit case, upon emission these photonic time-bin qudits will be entangled with the emitter. In general, the cyclic transitions commonly used for optical readout of various qudit systems can be used to optically pump photonic time-bin qudits with $q$ levels through a straightforward generalization of the pumping procedure described above for qubits. For example, consider a three-level quantum emitter with states $\ket{0}_\qd$, $\ket{1}_\qd$, $\ket{2}_\qd$. {We can initialize the emitter in the equal superposition state $\ket{0}_\qd+\ket{1}_\qd+\ket{2}_\qd$ and then alternate between pumping the system and performing an $X^{2}$ gate (Eq.~\eqref{eq:defXqudit}). Repeating this two-step cycle three times yields the state $\ket{0}_\qd\ket{0}_\p+\ket{1}_\qd\ket{1}_\p+\ket{2}_\qd\ket{2}_\p$, where $\ket{0}_\p$, $\ket{1}_\p$, $\ket{2}_\p$ are three different photonic time-bin states in chronological order. Denoting this net photon-pumping operation by ${\cal P}_\mathrm{pump}$, we have}
\begin{align}\label{eq:photon-pumping}
\begin{split}
&{\cal P}^{\mathrm{e},\mathrm{p}}_\mathrm{pump}(\ket{0}_\qd+\ket{1}_\qd+\ket{2}_\qd) = 
{\cal P}^{\mathrm{e},\mathrm{p}}_\mathrm{pump} \ \sum_{i=0}^2 \ket{i}_\qd \\
& = \ket{0}_\qd\ket{0}_\p+\ket{1}_\qd\ket{1}_\p+\ket{2}_\qd\ket{2}_\p \ ,
\end{split}
\end{align}
{where ${\cal P}^{\mathrm{e},\mathrm{p}}_\mathrm{pump}$ produces photon p from emitter e.
This can be generalized to qudits with arbitrarily many levels $q$:
\begin{align}\label{eq:photon-pumping-qudit}
\begin{split}
&{\cal P}^{\mathrm{e},\mathrm{p}}_\mathrm{pump}\sum_{i=0}^{\dloc-1}\, \ket{i}_\qd= \sum_{i=0}^{\dloc-1} \, \ket{i}_\qd\ket{i}_\p\\
& = \ket{0}_\qd\ket{0}_\p+\ket{1}_\qd\ket{1}_\p + \dots +\ket{\dloc -1}_\qd\ket{\dloc -1}_\p \ .
\end{split}
\end{align}
This state is obtained by applying $X^{q-1}$ to the emitter after each of the $q$ pumping operations. The full photon-pumping operation is depicted in Fig.~\ref{fig:photon-pumping}.
If we use the convention that photons that have not yet been emitted are initialized in state $\ket{0}_\mathrm{p}$, then we can express this operator as
\begin{equation}
    {\cal P}^{\mathrm{e},{\mathrm{p}}}_\mathrm{pump}=\sum_{i=0}^{q-1}\ket{i}_{\mathrm{e}}\bra{i}_{\mathrm{e}}\otimes X_{\mathrm{p}}^i \ .
\end{equation}
Note that we can modify the pumping operation to obtain different emitter-photon entangled states by changing the gates on the emitters. For example in the qutrit case described above, if we replace $X^{2}$ by $X$, then we obtain the state $\ket{0}_\qd\ket{0}_\p+\ket{1}_\qd\ket{2}_\p+\ket{2}_\qd\ket{1}_\p$. In the following section, we show that having access to these multiple versions of ${\cal P}_\mathrm{pump}$ is important for generating the various types of edges that qudit graph states can have. We will also see that ${\cal P}_\mathrm{pump}$ is the key ingredient needed for the deterministic generation of qudit graph states.
}

\section{Graph state generation}\label{sec:graph-state-generation}
 
{Any graph state of $n$ photonic qudits with $q$ levels each can be generated from a collection of coupled $q$-level quantum emitters using the operations described above. The simplest way to see this is to note that any photonic graph state can be generated in a conceptually straightforward manner by first preparing the target graph state on the emitters and then performing one photon-pumping operation on each emitter. If we then apply Hadamards on the emitters, measure the emitters, and finally apply measurement-adapted single-qudit gates on the photons, we can arrive at the target photonic graph state. 

To see how this works in more detail, we can first prepare the graph state on the emitters by preparing each emitter in the $\ket{X_0}$ state and then applying the gate $CZ_{\mathrm{e}_i,\mathrm{e}_j}^{\Gamma_{i,j}}$ on each pair of emitters $\mathrm{e}_i$ and $\mathrm{e}_j$, where $\Gamma$ is the adjacency matrix of the graph:
\begin{align}
&\left(\prod_{i<j}CZ_{\mathrm{e}_i,\mathrm{e}_j}^{\Gamma_{i,j}}\right)\ket{X_0}^{\otimes n}_\mathrm{e}\nonumber\\
&=\sum_{k_1,k_2,...,k_n=0}^{q-1}\left(\prod_{i<j}\omega^{k_ik_j\Gamma_{i,j}}\right)\ket{k_1k_2...k_n}_\mathrm{e} \ .
\end{align}
Because ${\cal P}^{\mathrm{e}_i,\mathrm{p}_i}_\mathrm{pump}$ and $CZ_{\mathrm{e}_i,\mathrm{e}_j}^{\Gamma_{i,j}}$ both leave the $Z$-basis states on emitter $\mathrm{e}_i$ invariant, it follows that these two operations commute. Therefore, the state after we apply the photon-pumping operation on each emitter can be rewritten as follows:
\begin{align}
&\left(\prod_{\ell=1}^n{\cal P}^{\mathrm{e}_\ell,\mathrm{p}_\ell}_\mathrm{pump}\right)\left(\prod_{i<j}CZ_{\mathrm{e}_i,\mathrm{e}_j}^{\Gamma_{i,j}}\right)\sum_{k_1,k_2,...,k_n=0}^{q-1}\ket{k_1k_2...k_n}_\mathrm{e}\nonumber\\ 
&=\left(\prod_{i<j}CZ_{\mathrm{e}_i,\mathrm{e}_j}^{\Gamma_{i,j}}\right)\left(\prod_{\ell=1}^n{\cal P}^{\mathrm{e}_\ell,\mathrm{p}_\ell}_\mathrm{pump}\right)\sum_{k_1,k_2,...,k_n=0}^{q-1}\ket{k_1k_2...k_n}_\mathrm{e}\nonumber\\
&=\left(\prod_{i<j}CZ_{\mathrm{e}_i,\mathrm{e}_j}^{\Gamma_{i,j}}\right)\sum_{k_1,k_2,...,k_n=0}^{q-1}\ket{k_1k_2...k_n}_\mathrm{e}\otimes\ket{k_1k_2...k_n}_\mathrm{p}\nonumber\\
&=\sum_{k_1,k_2,...,k_n=0}^{q-1}\left(\prod_{i<j}\omega^{k_ik_j\Gamma_{i,j}}\right)\ket{k_1k_2...k_n}_\mathrm{e}\otimes\ket{k_1k_2...k_n}_\mathrm{p}\nonumber\\
&=\left(\prod_{i<j}CZ_{\mathrm{p}_i,\mathrm{p}_j}^{\Gamma_{i,j}}\right)\sum_{k_1,k_2,...,k_n=0}^{q-1}\ket{k_1k_2...k_n}_\mathrm{e}\otimes\ket{k_1k_2...k_n}_\mathrm{p}\nonumber\\
&=\left(\prod_{i<j}CZ_{\mathrm{p}_i,\mathrm{p}_j}^{\Gamma_{i,j}}\right)\bigotimes_{i=1}^n\sum_{k_i=0}^{q-1}\ket{k_i}_{\mathrm{e}_i}\otimes\ket{k_i}_{\mathrm{p}_i}
\end{align}
If we now apply the Hadamard operation on each emitter and then measure it in the $Z$ basis, the state becomes
\begin{align}
&\left(\prod_{i<j}CZ_{\mathrm{p}_i,\mathrm{p}_j}^{\Gamma_{i,j}}\right)\bigotimes_{i=1}^n\ket{o_i}_{\mathrm{e}_i}\otimes\ket{X_{o_i}}_{\mathrm{p}_i} \ ,
\end{align}
where $o_i$ is the outcome from measuring emitter $\mathrm{e}_i$. If we then perform the operation $Z^{q-o_i}$ on each photon $\mathrm{p}_i$, and use that
\begin{equation}
    Z^{q-o_i}\ket{X_{o_i}}=\ket{X_{o_i+q-o_i\,\mathrm{mod}\,q}}=\ket{X_0},
\end{equation}
the state becomes
\begin{equation}
\left(\prod_{i<j}CZ_{\mathrm{p}_i,\mathrm{p}_j}^{\Gamma_{i,j}}\right)\ket{X_0}^{\otimes n}_\mathrm{p}\otimes\bigotimes_{i=1}^n\ket{o_i}_{\mathrm{e}_i} \ .
\end{equation}
We see that the final state of the photons is precisely the target graph state.

The above analysis shows that arbitrary graph states of $n$ photonic qudits can be generated using at most $n$ multi-level quantum emitters. However, for most (if not all) resource states of practical interest, far fewer than $n$ emitters are actually needed. This is because, in the above analysis, all the entanglement in the final resource state is created from emitter-emitter entangling gates. However, the emission process provides a second source of entanglement, since an emitted photon shares a maximally entangled link with the emitter it emerges from, as is evident from Eq.~\eqref{eq:photon-pumping-qudit}. If this entanglement can also be channeled into the target photonic graph state, then fewer emitters and emitter-emitter entangling gates will be needed. In this section, we show an extreme example of this, namely the 1D linear graph states. Figure~\ref{fig:qutrit-cluster} shows two small examples. Such states can be produced using only one emitter, regardless of how many photons they contain.
In Appendix~\ref{app:ladder}, we show how to construct a ladder state and more general 2D graph states referred to as cluster states (see Fig.~\ref{fig:ladder-and-2D}).
In particular, we show that a ladder state containing $2 \times n$ photons can be generated using two quantum emitters, while a more general 2D cluster state of size $m \times n$ with $m\le n$ can be produced from $m$ emitters. Unlike in the qubit case, in the case of qudits there are multiple types of 1D or 2D cluster states, depending on whether we use $CZ$ or $CZ^2$ or $CZ^3$, etc., for each edge. For example, two different linear graph states of three qudits are shown in Fig.~\ref{fig:qutrit-cluster}, where we use single edges to denote $CZ$ and double edges to denote $CZ^2$. We can of course also have graph states that contain both types of edges, triple edges, etc. 

\begin{figure} 
 \includegraphics[scale=0.19]{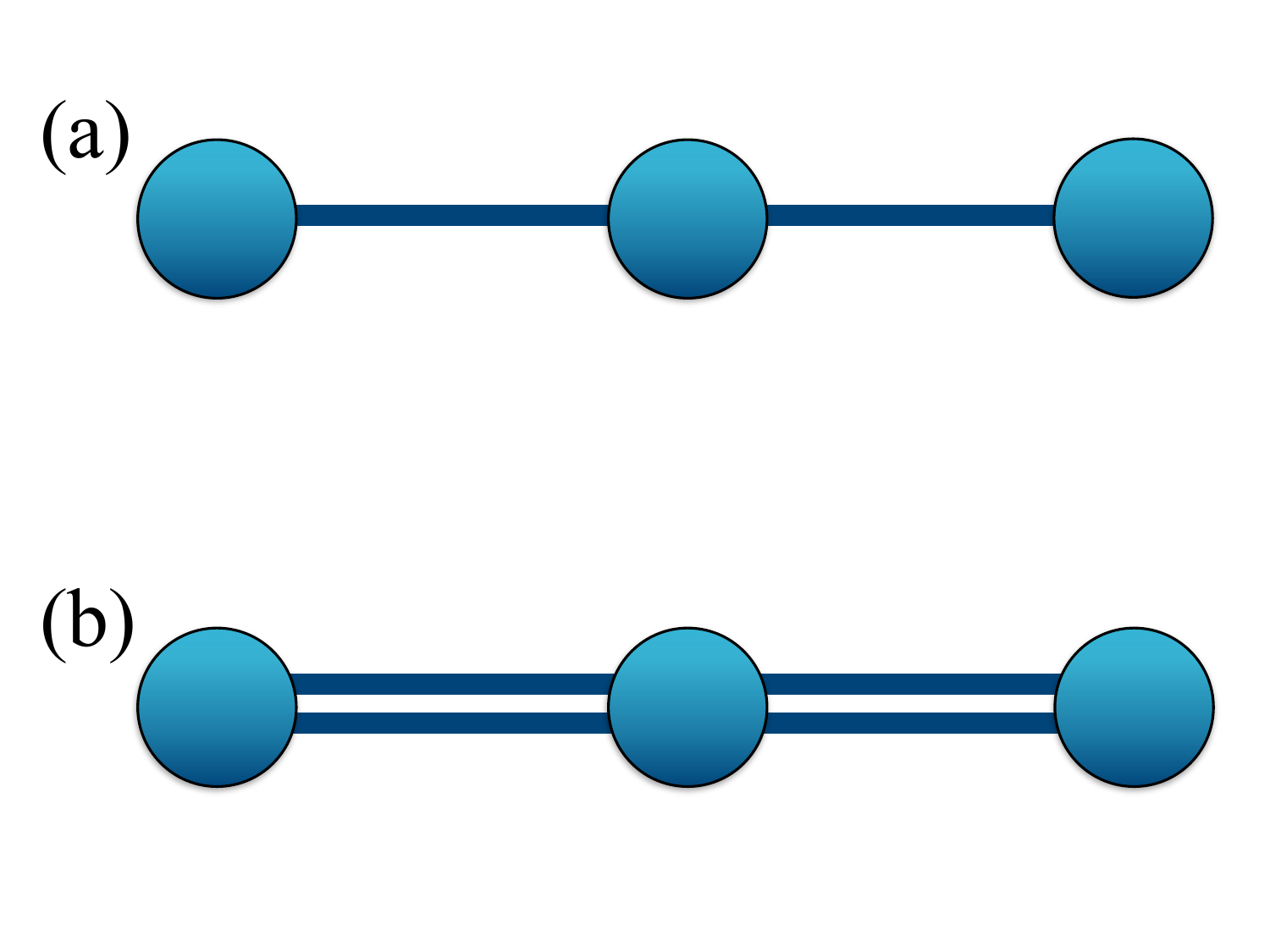} 
 \centering
 \caption{\label{fig:qutrit-cluster} 
  One-dimensional linear graph states of three qudits.
  (a) A linear graph state in which both edges correspond to $CZ$ gates.
  (b) A linear graph state in which both edges correspond to $CZ^2$ gates.
}
\end{figure}

  \begin{figure}
 \includegraphics[scale=0.29]{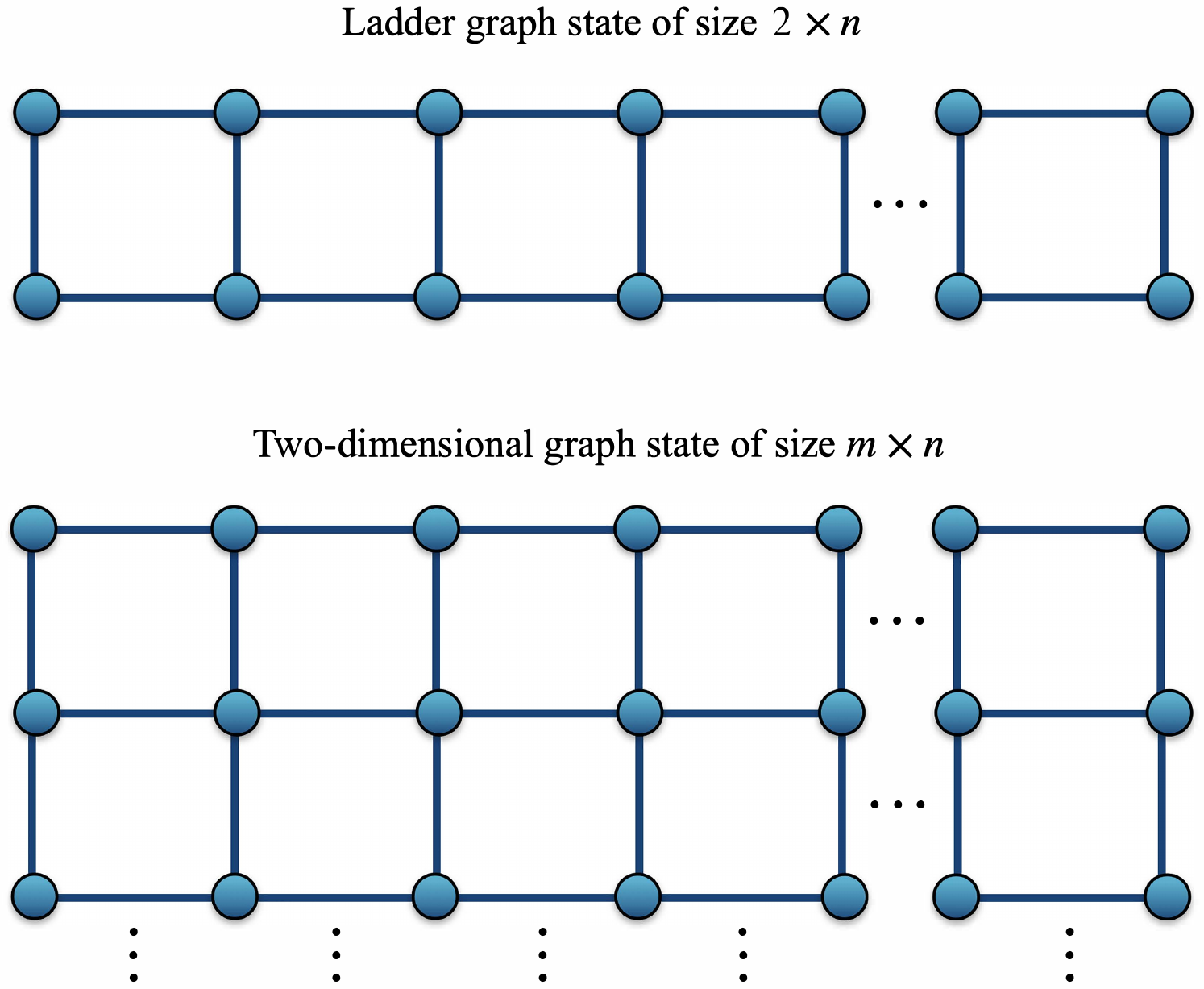}
 \centering
 \caption{\label{fig:ladder-and-2D} 
  {Two-dimensional linear graph states. (Top) Ladder state containing $2 \times n$ photons, and (bottom) 2D cluster state containing $m \times n$ photons.} }
\end{figure}
}

\begin{figure}
 \includegraphics[scale=0.4]{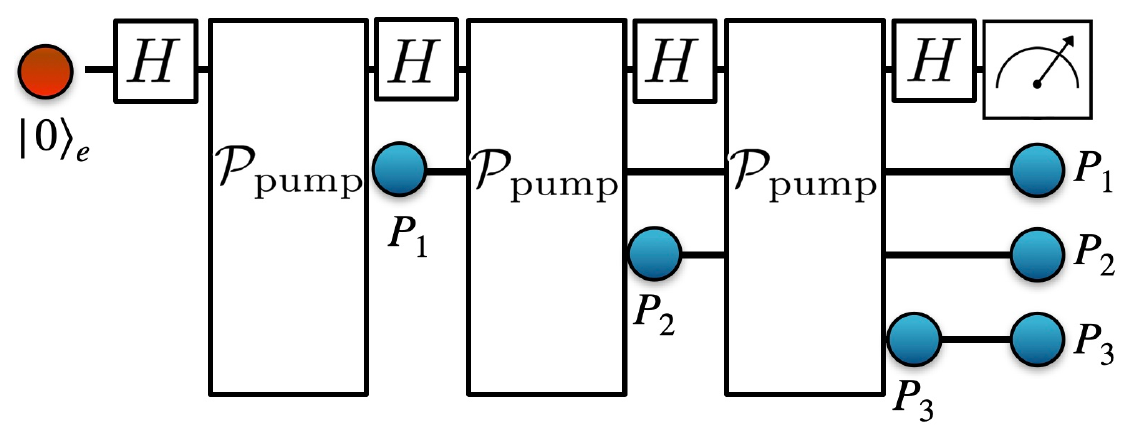} 
 \centering
 \caption{\label{fig:QC-qutrit-cluster} 
{Graph state generation circuit for 1D graph states of three photonic qudits like those shown in Fig.~\ref{fig:qutrit-cluster}.}
}
\end{figure}

{
We now show how to generate 1D linear photonic $q$-level qudit graph states comprised of only $CZ$ edges (as in Fig.~\ref{fig:qutrit-cluster}(a)) from a single $q$-level quantum emitter. Here, the only nonzero components in the upper triangle of the adjacency matrix are along the superdiagonal, $\Gamma_{i,i+1}=1$, and so the target graph state with $n$ photons is (using Eq.~\eqref{eq:general_graph_state})
\begin{equation}\label{eq:1D_cluster_state}
    \sum_{k_1,k_2,...,k_n=0}^{q-1}\left(\prod_{i=1}^{n-1}\omega^{k_ik_{i+1}}\right)\ket{k_1k_2...k_n}_\mathrm{p} \ .
\end{equation}
The main operations we need to produce this state are the qudit $H$ gate (Eq.~\eqref{eq:H-on-qudits}) and the photon-pumping operation ${\cal P}_\mathrm{pump}$ (Eq.~\eqref{eq:photon-pumping-qudit}). We first prepare the emitter in the state $\ket{0}_\mathrm{e}$ and then apply the $H$ gate to convert this to $\sum_{k_0=0}^{q-1}\ket{k_0}_\mathrm{e}$. Applying ${\cal P}_\mathrm{pump}$ to produce a photon yields the state $\sum_{k_0=0}^{q-1}\ket{k_0}_\qd\otimes\ket{k_0}_{\p_1}$, and performing a second $H$ gate on the emitter converts this to $\sum_{k_0,k_1=0}^{q-1} \omega^{k_0k_1} \ket{k_0}_\qd\otimes\ket{k_1}_{\p_1}$. We can then repeat the two-operation sequence $H{\cal P}_\mathrm{pump}$ an additional $n-1$ times to produce the following state of $n$ photons entangled with the emitter:
\begin{equation}
\sum_{k_0,k_1,...,k_n=0}^{q-1}\omega^{k_0k_1}\omega^{k_1k_2}...\omega^{k_{n-1}k_n}\ket{k_0}_\mathrm{e}\otimes\bigotimes_{\ell=1}^n\ket{k_\ell}_\mathrm{p_\ell} \ .
\end{equation}
We can then measure the emitter in the $Z$ basis to disentangle it from the photons. The post-measurement state is  
\begin{equation}
\ket{o_\qd}_\mathrm{e}\otimes\sum_{k_1,...,k_n=0}^{q-1}\omega^{o_\qd k_1}\omega^{k_1k_2}...\omega^{k_{n-1}k_n}\bigotimes_{\ell=1}^n\ket{k_\ell}_\mathrm{p_\ell} \ ,
\end{equation}
where $o_\qd$ is the measurement outcome. We can apply the operation $Z^{q-o_\qd}$ to photon $\p_1$ to finally bring the $n$-photon state into the form of the target state, Eq.~\eqref{eq:1D_cluster_state}:
\begin{equation}
\ket{o_\qd}_\mathrm{e}\otimes\sum_{k_1,...,k_n=0}^{q-1}\omega^{k_1k_2}...\omega^{k_{n-1}k_n}\bigotimes_{\ell=1}^n\ket{k_\ell}_\mathrm{p_\ell} \ .
\end{equation}
The full graph state generation circuit for the case of $n=3$ photons is depicted in Fig.~\ref{fig:QC-qutrit-cluster}.

The state shown in Fig.~\ref{fig:qutrit-cluster}(b) can be generated in a similar fashion, but now with a modified photon-pumping operation in which the $X^{q-1}$ gates on the emitters are replaced by $X^{q-2}$ gates. Starting from an equal superposition of emitter states, $\sum_{i=0}^{q-1}\ket{i}_\qd$, this modified version of ${\cal P}_\mathrm{pump}$ produces the state $\sum_{i=0}^{q-1}\ket{2i\hbox{ mod }q}_\qd\otimes\ket{i}_\p$. Applying an $H$ gate on the emitter then yields
\begin{equation}
	\sum_{i,j=0}^{q-1}\omega^{2ij}\ket{j}_\qd\otimes\ket{i}_\p \ .
\end{equation}
Comparing this to Eq.~\eqref{eq:CZqudit}, we see that the use of this modified ${\cal P}_\mathrm{pump}$ effectively generates a $CZ^2$ gate between emitter and photon instead of $CZ$. When the next photon is emitted, this becomes a $CZ^2$-type edge in the photonic graph state. More generally, we can create a $CZ^\beta$-type edge by using a modified ${\cal P}_\mathrm{pump}$ that uses $X^{q-\beta}$ gates on the emitters. In this way, we can not only create states like in Fig.~\ref{fig:qutrit-cluster}, but also states that contain multiple different edge types. We also note that there exists an alternative way to create $CZ^{q-1}$ edges where, instead of modifying ${\cal P}_\mathrm{pump}$, we can replace the $H$ gate by $H^\dag$ after each photon-pumping operation, as follows directly from Eq.~\eqref{eq:Hdag-on-qudits}. This alternative method will be used to generate various graph states in the following section.
}

{In the above examples, we create all the necessary entanglement via $CZ^\beta$ gates between emitters and via the emitter-photon entanglement that is created during the emission process ${\cal P}_\mathrm{pump}$. We do not make use of emitter-photon interaction in these examples. Such additional interaction after photon generation can be used to perform emitter-photon entangling gates like $CZ$, $CZ^2$, ... $CZ^{q-1}$. Allowing this third mechanism for entanglement generation can lead to even further reductions in the numbers of emitters needed to produce a target graph state. We give explicit examples of this in the next section, where we present protocols for generating AME states of photons. For example, in Sec.~\ref{sec:AME5q} and Appendix~\ref{app:AME5q}, we present and compare two methods of constructing an AME state of 5 qudits for any $q\ge 2$. In the first method, the use of additional interaction to perform emitter-photon CZ gates makes it possible to generate the entire state using only a single emitter. The second method does not assume interaction post photon generation is an option, in which case two emitters are needed to create the state. }

\section{AME state generation}\label{sec:AME-states}
Absolutely maximally entangled (AME) states, also referred to in some works as multipartite maximally  entangled states \cite{FacchiMES,Dardo2015,Zahra-Vahid}, are pure $n$-qudit quantum states with local dimension $\dloc$ such that every reduced density matrix on at most half the system size is maximally mixed.
For example, the Bell state $\ket{\phi^+} = \ket{00}+ \ket{11}$ and the GHZ state $\ket{\mathrm{GHZ}} = \ket{0000}+ \ket{1111}$ are both AME states because the reduced density matrices on each half of the system are
completely mixed. More formally, an $\AME(n,\dloc)$ state is a $n$-qudit pure state in $\hiH(n,\dloc)\coloneqq \mathbb{C}_\dloc^{\otimes n}$ iff
\begin{equation}
\rho_S = \Tr_{S^c} \ketbra\psi\psi \propto \1 \qquad \forall S \subset \{1,\ldots,n\}, |S| \leq \floor{n/2} \ , \nonumber
\end{equation}
where $S^c$ denotes the complementary set of $S$.
In this section, we describe how one can generate some of these states from quantum emitters.
{In the following, we use the notation $\ket{\AME(n,\dloc)}$ to refer to known AME states in their standard representation, whereas we use the notation $\ket{\AME(n,\dloc)_{\text{graph}}}$ to describe their graph state forms, which are local-unitary equivalent.}

\subsection{Generating AME states of $4$ qutrits}
While it is known that AME states of $4$ qubits do not exist \cite{Scott,Rains1999}, such states can exist for $\dloc >2$ \cite{Helwig2013b,Zahra-minimalsupport}. For example, in the case of 4 qutrits ($\dloc=3$), the corresponding AME state can be written explicitly as \cite{Helwig2013b}:
{
 \begin{align}\label{eq:ame43}
 \begin{split}
  \ket{\AME(4,3)} &= \sum_{\alpha,\beta=0}^2 \ket{\alpha , \beta , \alpha+ \beta \ \text{mod 3}, \alpha+2 \beta \ \text{mod 3}} \\
  & = \ket{0000} + \ket{0112} + \ket{0221} \\
  & +\ket{1011} + \ket{1120} + \ket{1202} \\
  & +\ket{2022} + \ket{2101} + \ket{2210}  \ ,
  \end{split}
 \end{align}
 where the local dimension is $q = 3$.  
 This state is locally equivalent to the graph states shown in Fig.~\ref{fig:AME4,3}  \cite{Helwig2013b}. In particular, one can perform $H$ gates on the 3rd and 4th qutrits of Eq.~\eqref{eq:ame43} to arrive at the graph form of the state: 
\begin{equation}\label{eq:graph-ame43}
 \ket{\AME(4,3)_{\text{graph}}} = \sum_{i, j , k, l = 0}^2 \omega^{i k} \omega^{j k} \omega^{i l} \, \omega^{2 j l} \,  \ket{i , j , k , l} \ .
\end{equation}
For the photon ordering shown in Fig.~\ref{fig:AME4,3}(a), this is
\begin{align}\label{eq:graph-ame43-ordering1}
 &\ket{\AME(4,3)_{\text{graph}}} = \nonumber\\ &\sum_{i, j , k, l = 0}^2 \omega^{i k} \omega^{j k} \omega^{i l} \, \omega^{2 j l} \,  \ket{i}_{\p_1}\ket{j}_{\p_4}\ket{k}_{\p_3}\ket{l}_{\p_2} \ .
\end{align}
while for the photon ordering shown in Fig.~\ref{fig:AME4,3}(b) we have
\begin{align}\label{eq:graph-ame43-ordering2}
 &\ket{\AME(4,3)_{\text{graph}}} = \nonumber\\ &\sum_{i, j , k, l = 0}^2 \omega^{i k} \omega^{j k} \omega^{i l} \, \omega^{2 j l} \,  \ket{i}_{\p_1}\ket{j}_{\p_4}\ket{k}_{\p_2}\ket{l}_{\p_3} \ .
\end{align}
}
\begin{figure}
 \includegraphics[scale=0.20]{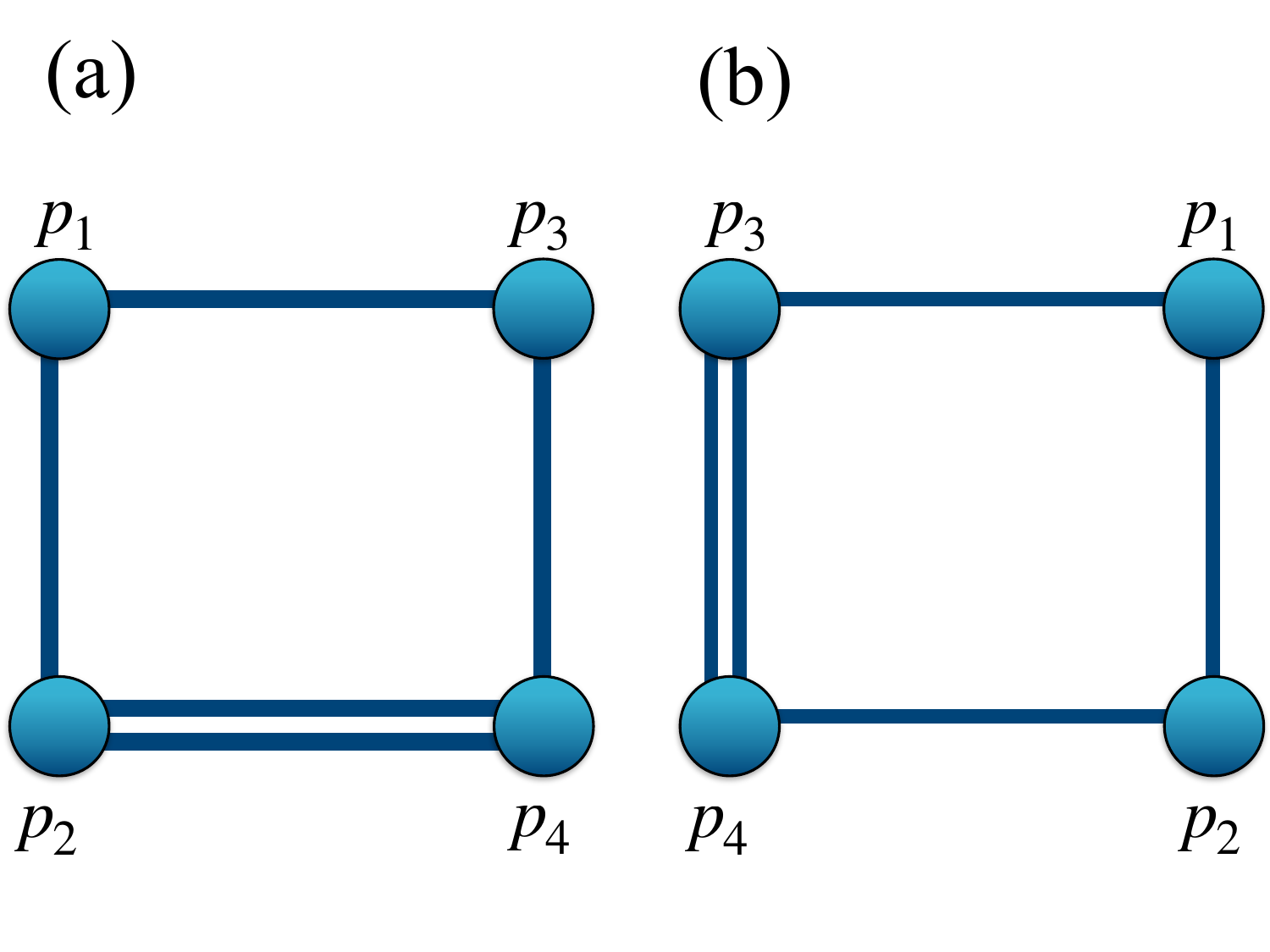}
 \centering
 \caption{\label{fig:AME4,3}
{AME state of $4$ qutrits.
  (a) The graph that represents the $\ket{\AME(4,3)_{\text{graph}}}$ state.
  (b) A graph representing a state that is the same as the one in (a) up to a rearrangement of the qutrits.}
}
\end{figure}

{
In what follows, we present protocols for generating the states in Eqs.~\eqref{eq:graph-ame43-ordering1} and \eqref{eq:graph-ame43-ordering2} from two quantum emitters, each of which has three ground levels, one of which ($\ket{0}_\qd$) is optically coupled to an excited state via a cyclic transition. Although the two states in Eqs.~\eqref{eq:graph-ame43-ordering1} and \eqref{eq:graph-ame43-ordering2} are the same up to a swapping of two photonic qutrits, there are important physical distinctions here. First, the order in which the photons are produced is different in the two cases, as we describe below; this can be an important factor in how the resource state is used, since it is generally preferable to produce photons in the order in which they are later measured to reduce photon storage requirements~\cite{Li2022}. In addition, the two protocols require different types of emitter gates.  In both protocols, we make use of entangling operations between the emitters, in the spirit of Refs.~\cite{Sophia-2Dcluster,ButerakosPRX2017,
 Russo_2emitters,Segovia-2emitters,
Michaels-2emitters,Hilaire2021resource,Li2022}. However, method 1 requires the ability to perform $H$, $H^\dag$, and $CZ$ on the emitters, while method 2 requires $H$, $CZ$, and $CZ^2$. Thus, one can trade one type of entangling gate for an additional type of single-emitter gate in this case. The quantum circuits for both protocols are presented in Fig.~\ref{fig-QC-AME4,3}.}

\begin{figure}
 \includegraphics[scale=0.55]{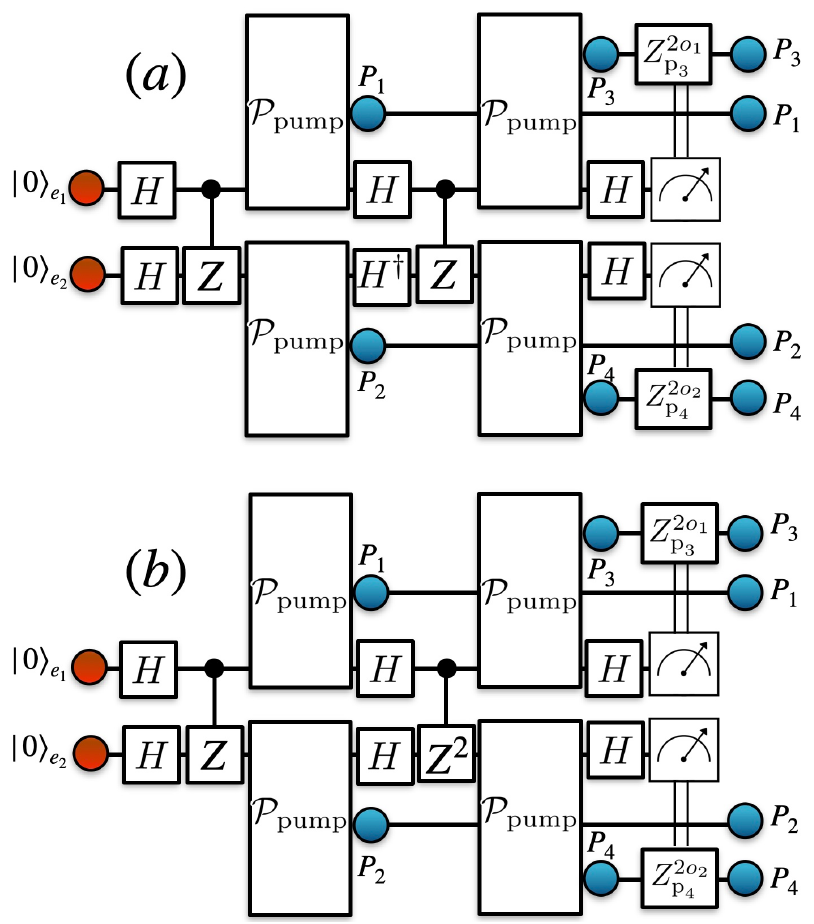}
 \centering
 \caption{\label{fig-QC-AME4,3}
{Quantum circuits that produce the two versions of the state $\ket{\AME(4,3)_{\text{graph}}}$ shown in Fig.~\ref{fig:AME4,3}.}
}
\end{figure}

{
Our first method of generating an AME state of $4$ qutrits (Fig.~\ref{fig-QC-AME4,3}(a)) involves the following steps:
\begin{itemize}[leftmargin=.5in]
   \item[Step 1.] Prepare the two emitters in the state \\
   $\hookrightarrow \ket{\phi\step1} = \ket{0}_{\qd_1}\ket{0}_{\qd_2}$   
   
   \item[Step 2.] $H$ gate, Eq.~\eqref{eq:H-on-qudits}, on each emitter\\
   $\hookrightarrow \ket{\phi\step2}  =  \sum_{i,l=0}^2 \ket{i}_{\qd_1}\ket{l}_{\qd_2}$
   
   \item[Step 3.] $CZ$ gate, Eq.~\eqref{eq:CZqutrit}, on the two emitters\\
    $\hookrightarrow \ket{\phi\step3}  =\sum_{i,l=0}^2 \omega^{il} \  \ket{i}_{\qd_1}\ket{l}_{\qd_2}$
   
   \item[Step 4.] ${\cal P}_\mathrm{pump}$, Eq.~\eqref{eq:photon-pumping}, on each emitter\\
   $\hookrightarrow \ket{\phi\step4}  =\sum_{i,l=0}^2 \omega^{il} \  \ket{i}_{\qd_1} \ket{i}_{\p_1} \ket{l}_{\qd_2} \ket{l}_{\p_2}$
   
   \item[Step 5.] $H$ gate on the 1st and $H^\dag$ gate, Eq.~\eqref{eq:Hdag-on-qudits}, on the 2nd emitter\\
   $\hookrightarrow \ket{\phi\step5} = \\ \sum_{i,j,k,l=0}^2 \omega^{il}  \omega^{ik} \omega^{2jl}\  \ket{k}_{\qd_1} \ket{i}_{\p_1} \ket{j}_{\qd_2} \ket{l}_{\p_2}$
   
   \item[Step 6.] $CZ$ gate on the two emitters\\
   $\hookrightarrow \ket{\phi\step6}  =\\ \sum_{i,j,k,l=0}^2  \omega^{ik}  \omega^{jk} \omega^{il} \omega^{2jl} \ \ket{k}_{\qd_1} \ket{i}_{\p_1} \ket{j}_{\qd_2} \ket{l}_{\p_2}$
   
   \item[Step 7.] ${\cal P}_\mathrm{pump}$ on each emitter\\
   $\hookrightarrow \ket{\phi\step7}  =\\\sum_{\substack{i,j,\\k,l=0}}^2 \omega^{ik}  \omega^{jk} \omega^{il} \omega^{2jl} \  \ket{k}_{\qd_1}\ket{k}_{\p_3} \ket{i}_{\p_1} \ket{j}_{\qd_2}\ket{j}_{\p_4} \ket{l}_{\p_2}$
   
      \item[Step 8.] $H$ gate on each emitter and measure each emitter in the $Z$ basis \\
   $\hookrightarrow \ket{\phi\step8}  =\ket{\AME(4,3)_{\text{graph}}}$\\$=\sum_{i,j,k,l=0}^2   \omega^{ik}  \omega^{jk} \omega^{il} \omega^{2jl} \   \ket{i}_{\p_1} \ket{j}_{\p_4} \ket{k}_{\p_3}\ket{l}_{\p_2}$
 \end{itemize}
 }
{To yield the state in the last step, we also performed the local gates $Z_{\p_3}^{2o_1}$, $Z_{\p_4}^{2o_2}$ on photons 3 and 4, where $o_1$ and $o_2$ are the measurement outcomes for emitters 1 and 2, respectively. The final state in step 8 is the state $\ket{\AME(4,3)_{\text{graph}}}$, shown in Eq.~\eqref{eq:graph-ame43-ordering1}. All the steps are summarized by the quantum circuit shown in Fig.~\ref{fig-QC-AME4,3}(a).
}

\begin{figure*}
 \includegraphics[width=0.8\textwidth]{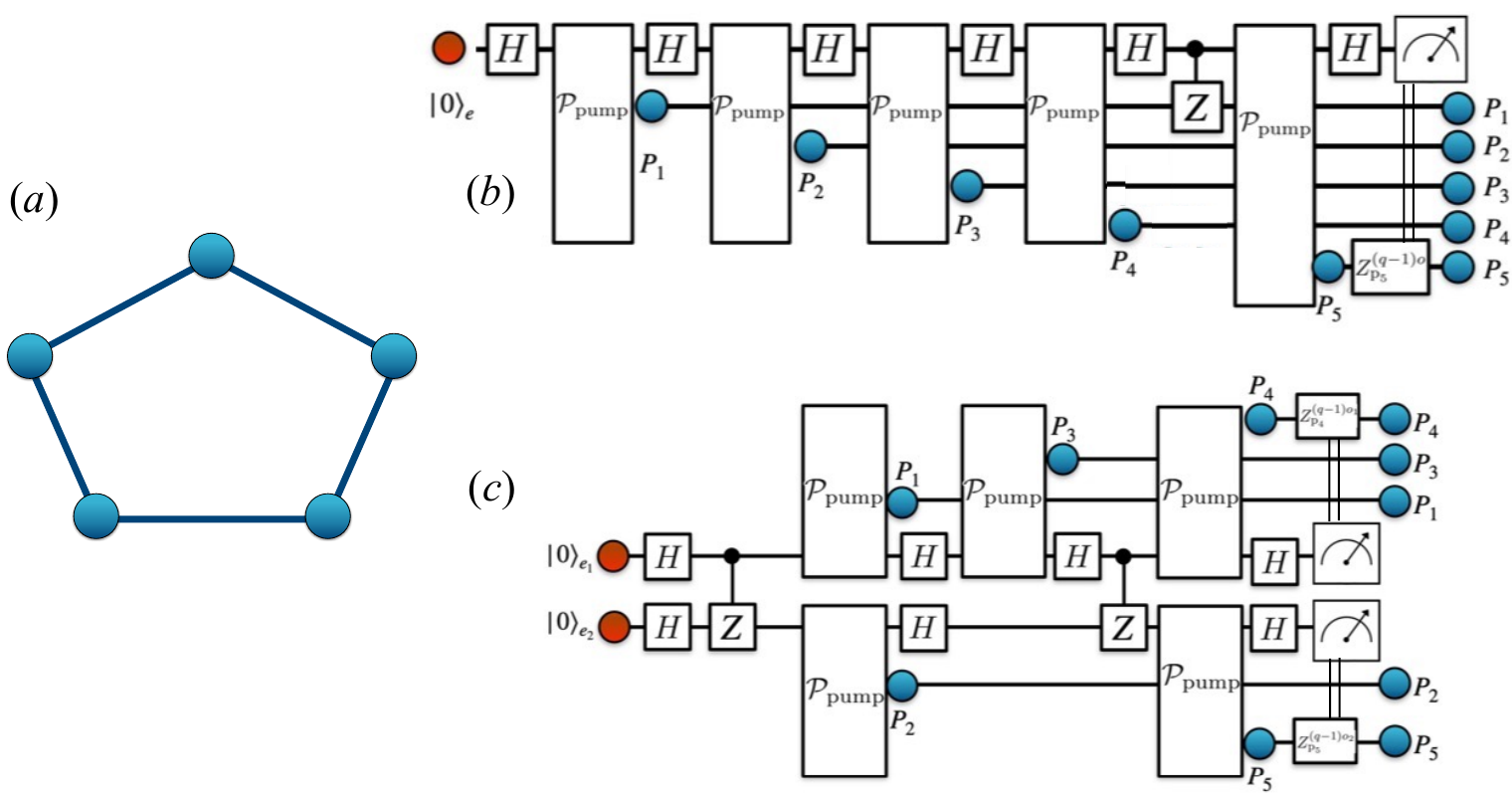}
 \centering
 \caption{\label{fig:AME5,q} {
(a) Graph representation of the $\ket{\AME(5,q)_{\text{graph}}}$ state. (b) Generation circuit for the state in (a) that uses only a single quantum emitter with $\dloc$ ground states. This circuit assumes the use of emitter-photon interaction. (c) Generation circuit for the state in (a) that uses two quantum emitters, each with $\dloc$ ground states. This circuit does not require emitter-photon interaction.}
}
\end{figure*}

To produce the state shown in Fig.~\ref{fig:AME4,3}(b), we propose the following slightly modified protocol. The first 4 steps remain the same as above, but steps 5-8 are different:
{
 \begin{itemize}[leftmargin=.5in]
   \item[Step 5'.] $H$ gate, Eq.~\eqref{eq:H-on-qudits}, on each emitter\\
    $\hookrightarrow \ket{\phi\step5}  = \\\sum_{i,j,k,l=0}^2 \omega^{il}  \omega^{ik} \omega^{jl}\  \ket{k}_{\qd_1} \ket{i}_{\p_1} \ket{j}_{\qd_2} \ket{l}_{\p_2}$
    
      \item[Step 6'.] $CZ^2$ gate, Eq.~\eqref{eq:CZ2qutrit}, on the two emitters\\
      $\hookrightarrow \ket{\phi\step6}  $\\
      $=\sum_{i,j,k,l=0}^2  \omega^{il}  \omega^{ik} \omega^{jl} \omega^{2jk} \  \ket{k}_{\qd_1} \ket{i}_{\p_1} \ket{j}_{\qd_2} \ket{l}_{\p_2}$ 
      
   \item[Step 7'.] ${\cal P}_\mathrm{pump}$, Eq.~\eqref{eq:photon-pumping}, on each emitter\\
    $\hookrightarrow \ket{\phi\step7} =$\\
    $\sum_{\substack{i,j,\\k,l=0}}^2\omega^{il}  \omega^{ik} \omega^{jl} \omega^{2jk}\  \ket{k}_{\qd_1}\ket{k}_{\p_3} \ket{i}_{\p_1} \ket{j}_{\qd_2}\ket{j}_{\p_4} \ket{l}_{\p_2}$
   
   \item[Step 8'.] $H$ gate on each emitter\\
       $\hookrightarrow \ket{\phi\step8} =\\\sum_{\substack{i,j,k,\\l,m,n=0}}^2 \Theta \  \ket{m}_{\qd_1}\ket{k}_{\p_3} \ket{i}_{\p_1} \ket{n}_{\qd_2}\ket{j}_{\p_4} \ket{l}_{\p_2}$,\\
       where $\Theta \coloneqq  \omega^{il}  \omega^{ik} \omega^{jl} \omega^{2jk} \omega^{km} \omega^{jn} $
   
    \item[Step 9'.] Measure each emitter in the $Z$ basis\\
   $\hookrightarrow \ket{\phi\step9} =\ket{\AME(4,3)_\text{graph}}$
   \\$=\sum_{i,j,k,l=0}^2  \omega^{ik}  \omega^{jl} \omega^{il}   \omega^{2jk} \   \ket{i}_{\p_1} \ket{j}_{\p_4} \ket{k}_{\p_3} \ket{l}_{\p_2}$
 \end{itemize}
As before, after the emitters are measured, we perform the local gates $Z_{\p_3}^{2o_1}Z_{\p_4}^{2o_2}$ on photons 3 and 4, where $o_1$ and $o_2$ are the measurement outcomes for emitters 1 and 2, respectively, to arrive at the final state. This state is the same as in Eq.~\eqref{eq:graph-ame43-ordering2} with the indices $k\leftrightarrow l$ relabeled. The quantum circuit that produces this state is shown in Fig.~\ref{fig-QC-AME4,3}(b). Note also that larger ladder-type qudit graph states with arbitrary combinations of $CZ$ and $CZ^2$ edges can be produced by repeating steps 5-7 in the above protocols. This is discussed in further detail in Appendix~\ref{app:ladder}.}

Comparing the circuits in Figs.~\ref{fig-QC-AME4,3}(a) and (b), we see that we have the freedom to choose between performing a $CZ^2$ gate on the emitters or performing an $H^\dag$ (instead of $H$) gate on one emitter followed by a $CZ$ gate on the two emitters. {However, these two circuits are not physically equivalent since the vertices of the graph are generated in a different order in the two cases. In both protocols, photons $\p_1$ and $\p_2$ are emitted first, followed by $\p_3$ and $\p_4$. However, $\p_2$ and $\p_3$ are swapped in the corresponding graphs (compare the two graphs in Fig.~\ref{fig:AME4,3}), and these two vertices are not equivalent to each other. Thus, the order in which the photons will later be measured may determine which protocol is the better option.}

\subsection{Generating AME states of 5 qudits}\label{sec:AME5q}

AME states of $5$ qubits were shown to exist in Ref.~\cite{Laflamme1996}. 
By exploiting a connection to quantum orthogonal arrays, explicit examples of $\AME(5,q)$ states for any local dimension $q \geq 2$ were discovered more recently \cite{Zahra-QOA}:
{ \begin{equation}\label{eq:ame5q} 
  \ket{\AME(5,\dloc)} = \sum_{\alpha, \beta=0}^{\dloc -1} \ket{ \alpha,\beta,\alpha+\beta \ \text{mod}\ q} \ket{\phi_{\alpha,\beta}}, 
 \end{equation}}
where $\ket{\phi_{\alpha,\beta}} = X^\alpha \otimes Z^\beta \sum_{\gamma=0}^{\dloc -1} \ket{\gamma , \gamma}$.
One can also show that this state has a graph representation.
In particular, if one performs Hadamard gates on the third and fourth qudits, the resulting state is
{
\begin{align}
\begin{split}
&\ket{\AME(5,\dloc)_{\text{graph}}} \\
&=\sum_{i,j,k,l,m=0}^{q-1} \omega^{ij} \omega^{jk} \omega^{kl} \omega^{lm} \omega^{mi} \, \ket{i,j,k,l,m} \ ,
\end{split}
\end{align}
which is the graph state shown in Fig.~\ref{fig:AME5,q}(a).}

Here, we present two general protocols for generating such states that work for all values of the local dimension $\dloc$. 
 The first protocol is summarized in Fig.~\ref{fig:AME5,q}(b). The state after each layer of the circuit is shown in Appendix~\ref{app:AME5q}. This protocol uses only a single quantum emitter, but it assumes the capability of an emitted photon to interact again with the emitter. 
{More specifically, it requires the ability to perform one emitter-photon $CZ$ gate between one of the emitted photons and the emitter.
While several works have proposed schemes for doing this in the case of photonic qubits \cite{Pichler2017,Zhan_PRL2020,Kianna-feedback,Shi-feedback}, the analogous qudit operation may be substantially more challenging. 
This motivates the development of an alternative protocol that does not require such an operation. 
}

\begin{figure*}
 \includegraphics[width=0.8\textwidth]{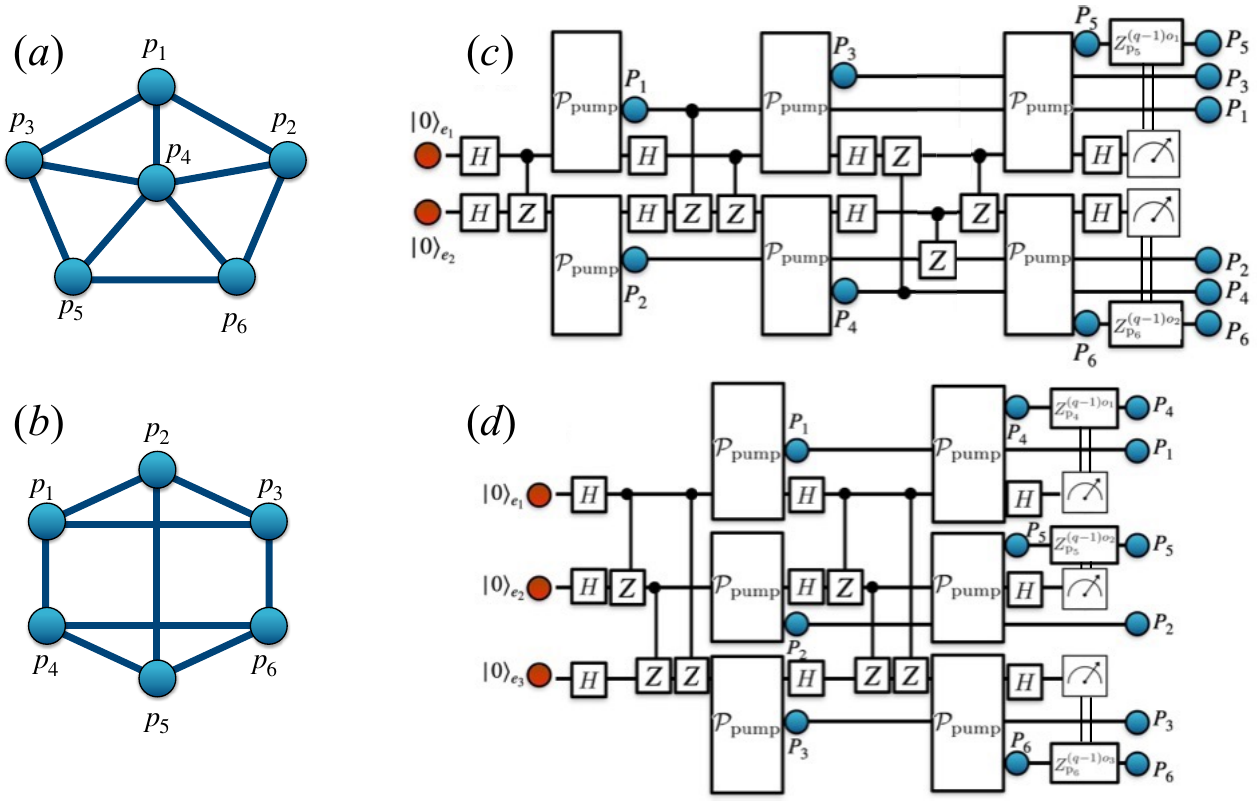}
 \centering
 \caption{\label{fig:AME6,q} {
(a,b) Two types of $\AME(6,\dloc)$ states represented as graphs. (c) Circuit that generates the state in (a) using two quantum emitters and emitter-photon interaction. (d) Circuit that generates the state in (b) using three quantum emitters but no emitter-photon interaction.} 
}
\end{figure*}
\begin{figure*}
 \includegraphics[width=0.9\textwidth]{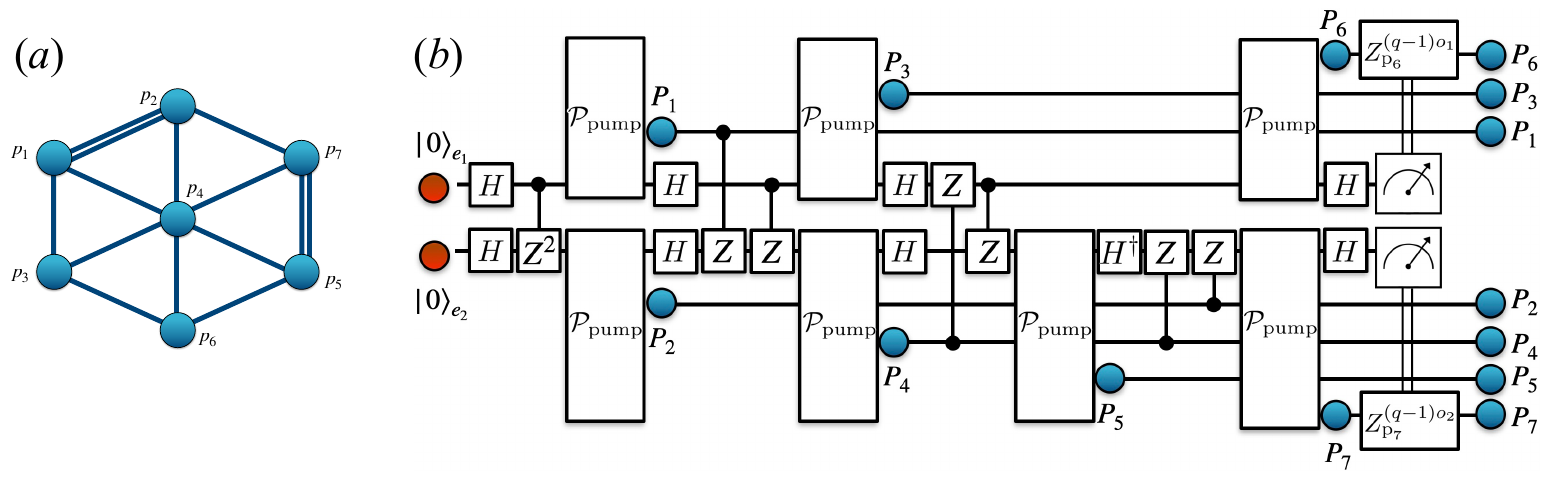}
 \centering
 \caption{\label{fig:AME7,3} {
(a) Graph representing the $\ket{\AME(7,3)_{\text{graph}}}$ state. (b) Circuit that generates the state in (a) using two emitters and photon-emitter interaction.}
}
\end{figure*}

 The second protocol does not require photon-emitter interaction, but at the expense of needing two coupled quantum emitters. {This protocol is summarized in Fig.~\ref{fig:AME5,q}(c), where it is evident that two emitter-emitter $CZ$ gates are used. Aside from these gates and the photon-pumping operation, the protocol only requires $H$ gates on the emitters and $Z^\beta$ gates on some photons. The state after each layer of the circuit is shown in Appendix~\ref{app:AME5q}. In both protocols, we assume the emitters have $\dloc$ ground states, one of which can be optically excited to a higher-energy state via a cyclic transition. Both use the photon-pumping operation defined in Eq.~\eqref{eq:photon-pumping}.}

\subsection{Generating AME states of $6$ qudits}

Next, we consider $\ket{\AME(6,\dloc)_{\text{graph}}}$ states. The graph representations of two such states are shown in Fig.~\ref{fig:AME6,q} \cite{Helwig2013b,Alba-qudit-graphstates}.
We present protocols for generating both of these types of graph states for qudits of arbitrary local dimension $\dloc$.
The first protocol utilizes two coupled emitters with $\dloc$ ground levels each and photon-emitter interaction to generate graph states of the sort shown in Fig.~\ref{fig:AME6,q}(a).
{The generation circuit is shown in Fig.~\ref{fig:AME6,q}(c).
The gates we use in this protocol are  $H$ gates (Eq.~\ref{eq:H-on-qudits}) and $CZ$ (Eq.~\ref{eq:CZqudit}) gates on the two emitters. This protocol also requires three emitter-photon $CZ$ gates generated by interaction after the photons are generated. The state after each layer of the circuit is given in Appendix~\ref{app:AME6q}. 
} 

{Next, we present a protocol for generating the $\AME(6,\dloc)$ state corresponding to the graph shown in Fig.~\ref{fig:AME6,q}(b). This protocol requires three coupled emitters with $\dloc$ ground levels each but does not need photon-emitter interaction. The circuit is shown in Fig.~\ref{fig:AME6,q}(d). The only gates it requires are $H$ and $CZ$ gates on emitters, as well as $Z^\beta$ gates on the photons. The state after each layer of the circuit is given in Appendix~\ref{app:AME6q}. 
}

\subsection{Generating AME states of $7$ qutrits}
It is known that while AME states of $7$ qubits do not exist \cite{Huber-AME7qubits}, AME states of $7$ qutrits do exist.
Through an exhaustive numerical search checking the bipartite entanglement of various graph states, it was found that the graph shown in Fig.~\ref{fig:AME7,3}(a) corresponds to the $\AME(7,3)$ state \cite{Helwig2013b}. {We found that it is possible to generate this state using two quantum emitters and emitter-photon interaction. The circuit is shown in Fig.~\ref{fig:AME7,3}(b), while the state after each layer of the circuit can be found in 
Appendix~\ref{app:AME73}. 
Our protocol uses two quantum emitters with three ground levels each, as well as photon-emitter interaction (four $CZ$ operators between photons and emitters). }
The protocol is similar to the one presented above for the AME(6,$\dloc$) states depicted in Fig.~\ref{fig:AME6,q}(a) due to the similarity in graph structure.

 \begin{widetext}

\begin{table}[t]
\begin{center}
\begin{tabular}{ |c|c|c|c|c|}  
 \hline
 Graph & \# photons & Local dimension & \# emitters & Photon interaction required? \\
 \hline
  \hline
           \parbox[c]{16em}{\includegraphics[width=1.6in]{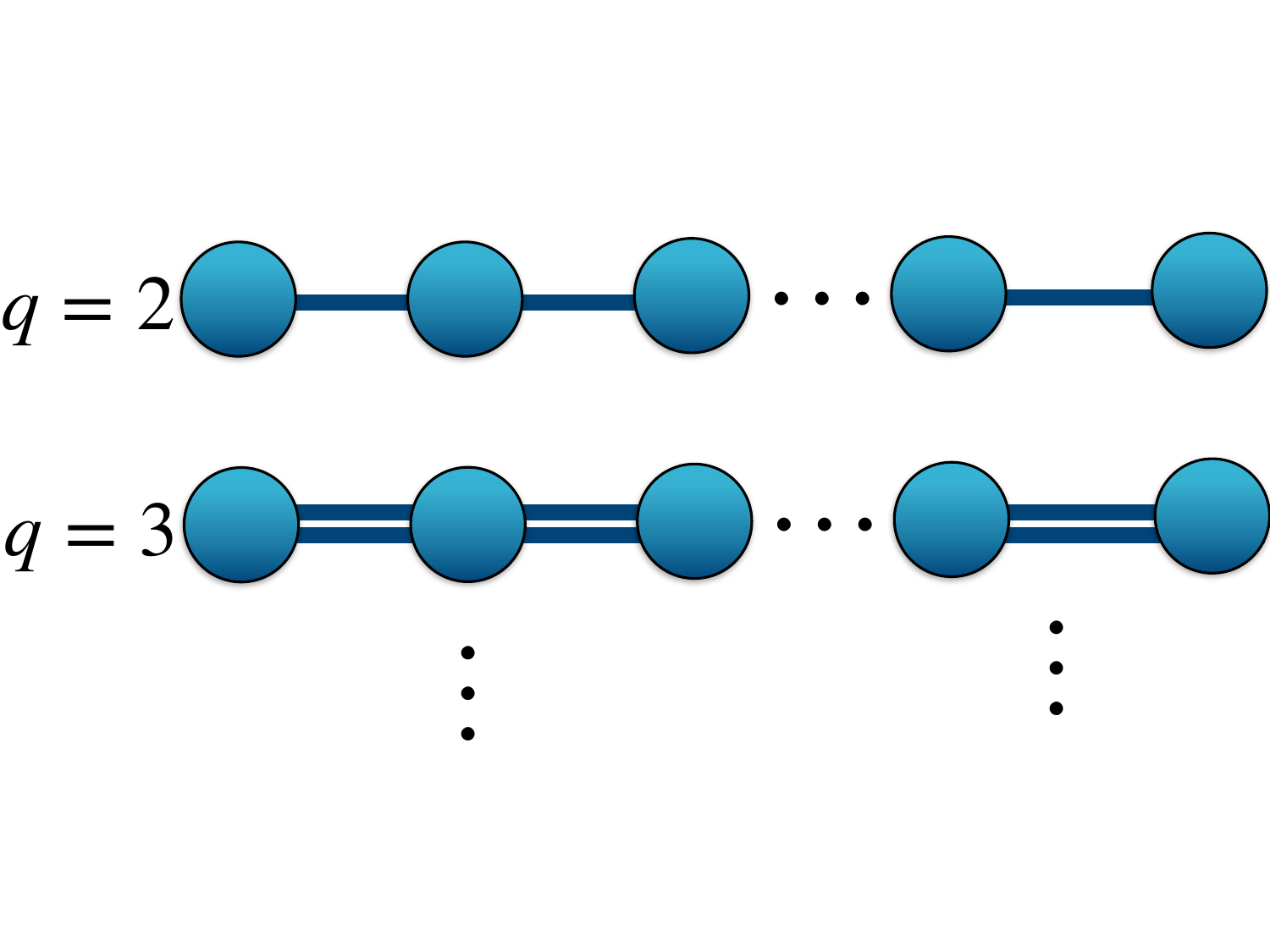}} & 
 $n \geq 2$ &  
 $\dloc \geq 2$ &
 1 &  
 no
 \vspace*{-5mm}
         \\ 
 \hline
             \parbox[c]{16em}{\includegraphics[width=1.6in]{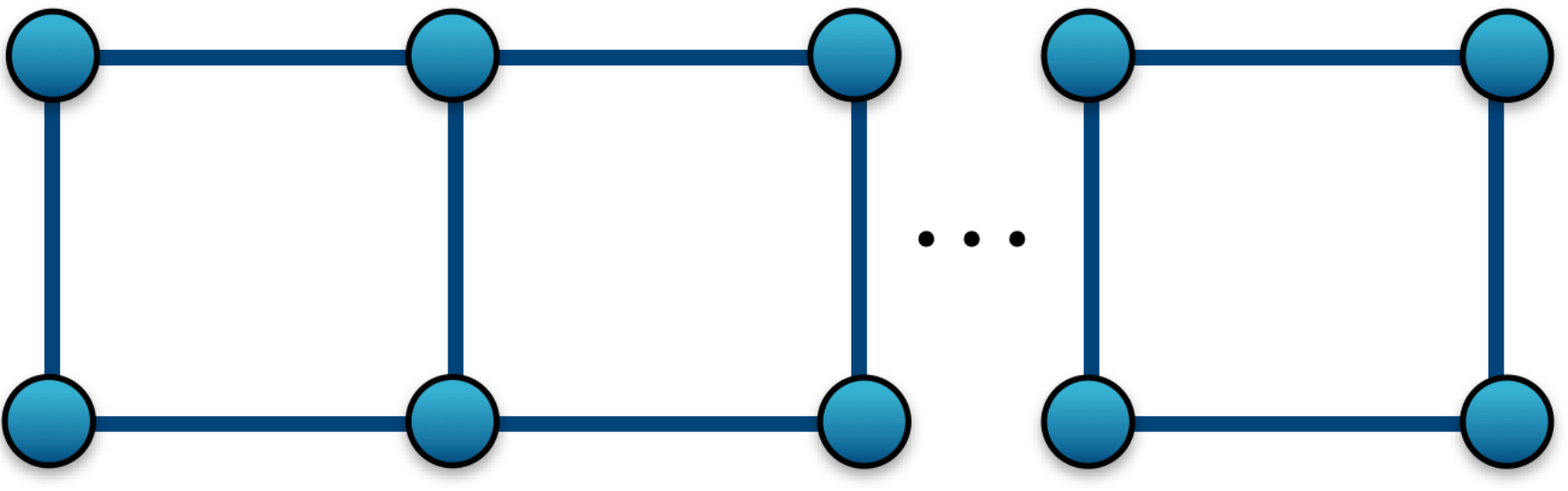}} & 
 $n \geq 4$ &  
 $\dloc \geq 2$ &
 2 &  
 no
          \\ 
 \hline
             \parbox[c]{16em}{\includegraphics[width=1.5in]{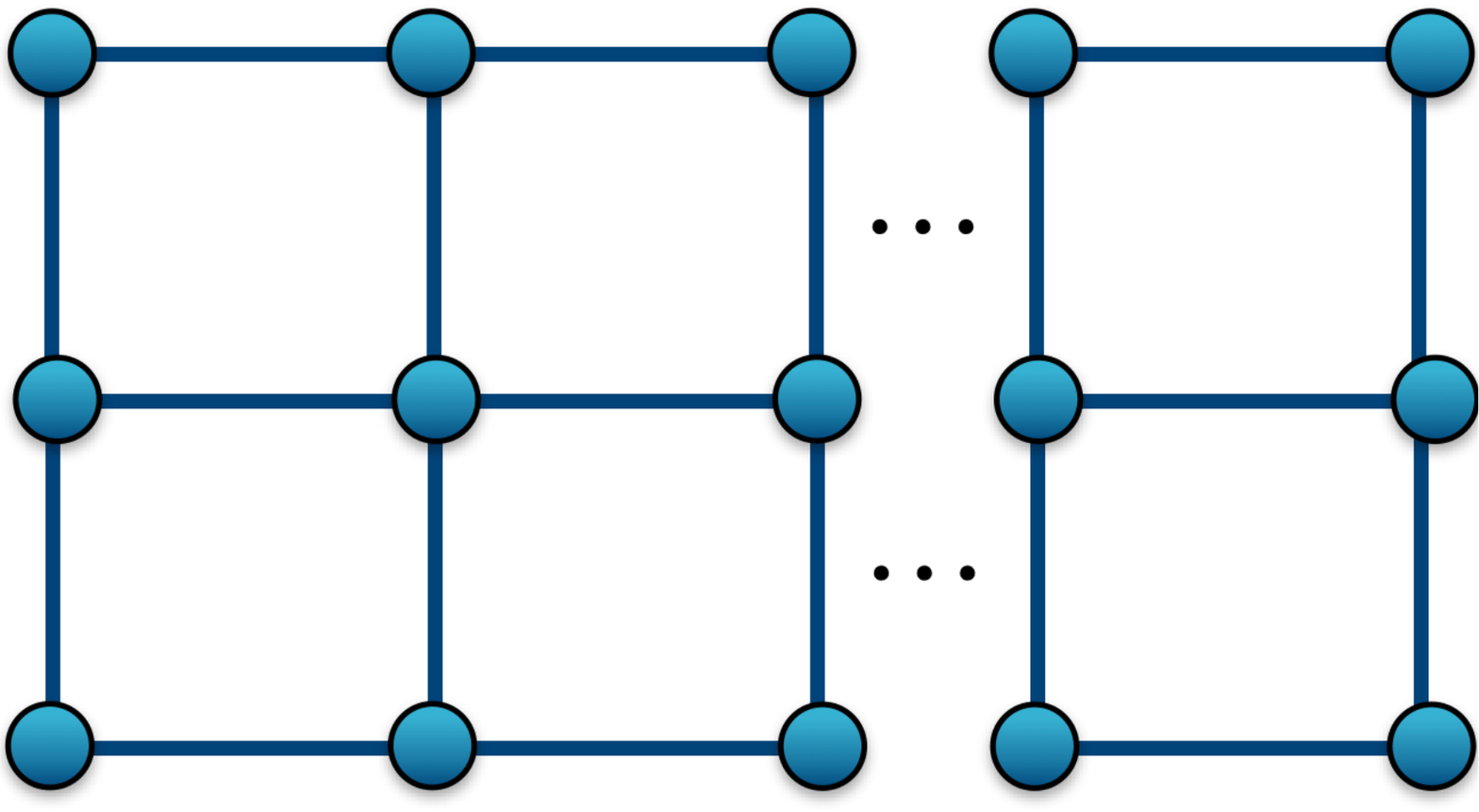}} & 
 $n \geq 6$ &  
 $\dloc $ &
 3 &  
 no
         \\ 
  \hline
            \parbox[c]{16em}{\includegraphics[width=1.3in]{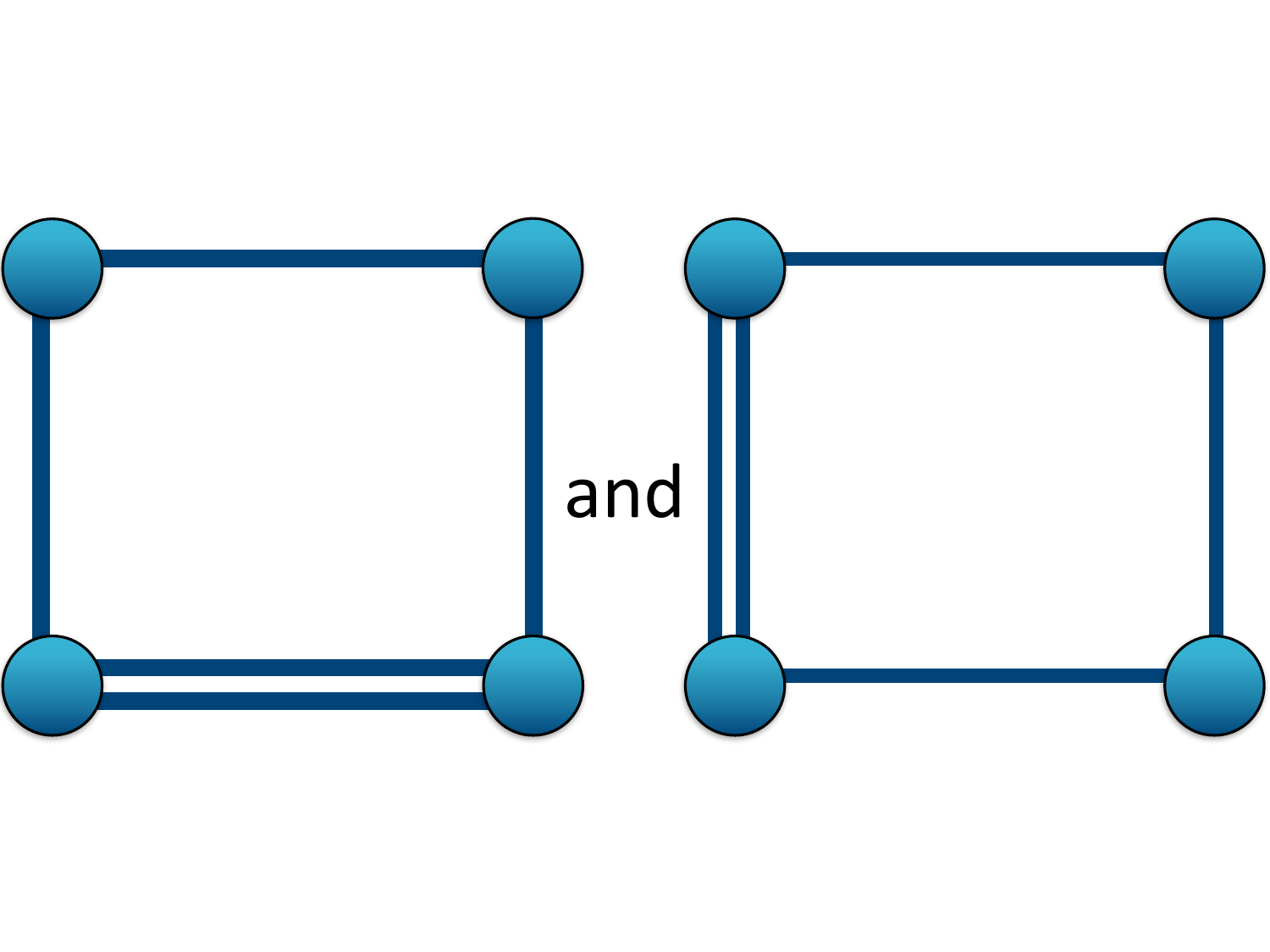}} & 
 $n =4$ &  
 $\dloc =3$ &
 2 &  
 no
         \\ 
  \hline
            \parbox[c]{16em}{\includegraphics[width=0.85in]{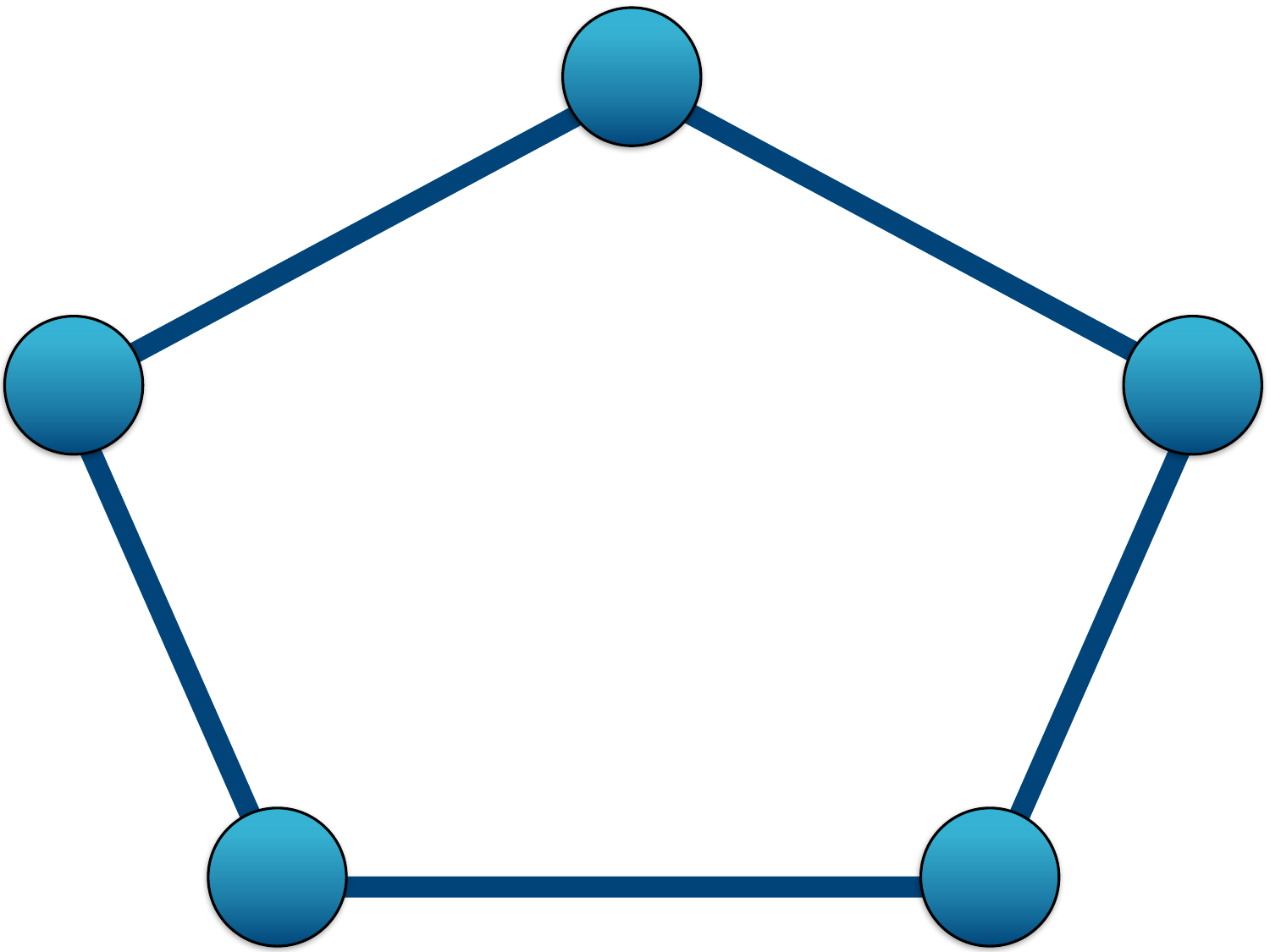}} & 
 $n =5$ &  
 $\dloc \geq 2$ &
 $ \begin{tabular}{c}
    \text{First method: 1  } \\

    \rule{3cm}{0.4pt}
   \vspace{0.1cm}
  \\  
    
    \text{Second method: 2  }  
 \end{tabular} $ & 
  $ \begin{tabular}{c}
    \text{yes}  \\ 
    
\rule{4cm}{0.4pt}
 \vspace{0.1cm}
\\
    \text{no}  \\
 \end{tabular} $  
      \\ 
 \hline
             \parbox[c]{16em}{\includegraphics[width=0.85in]{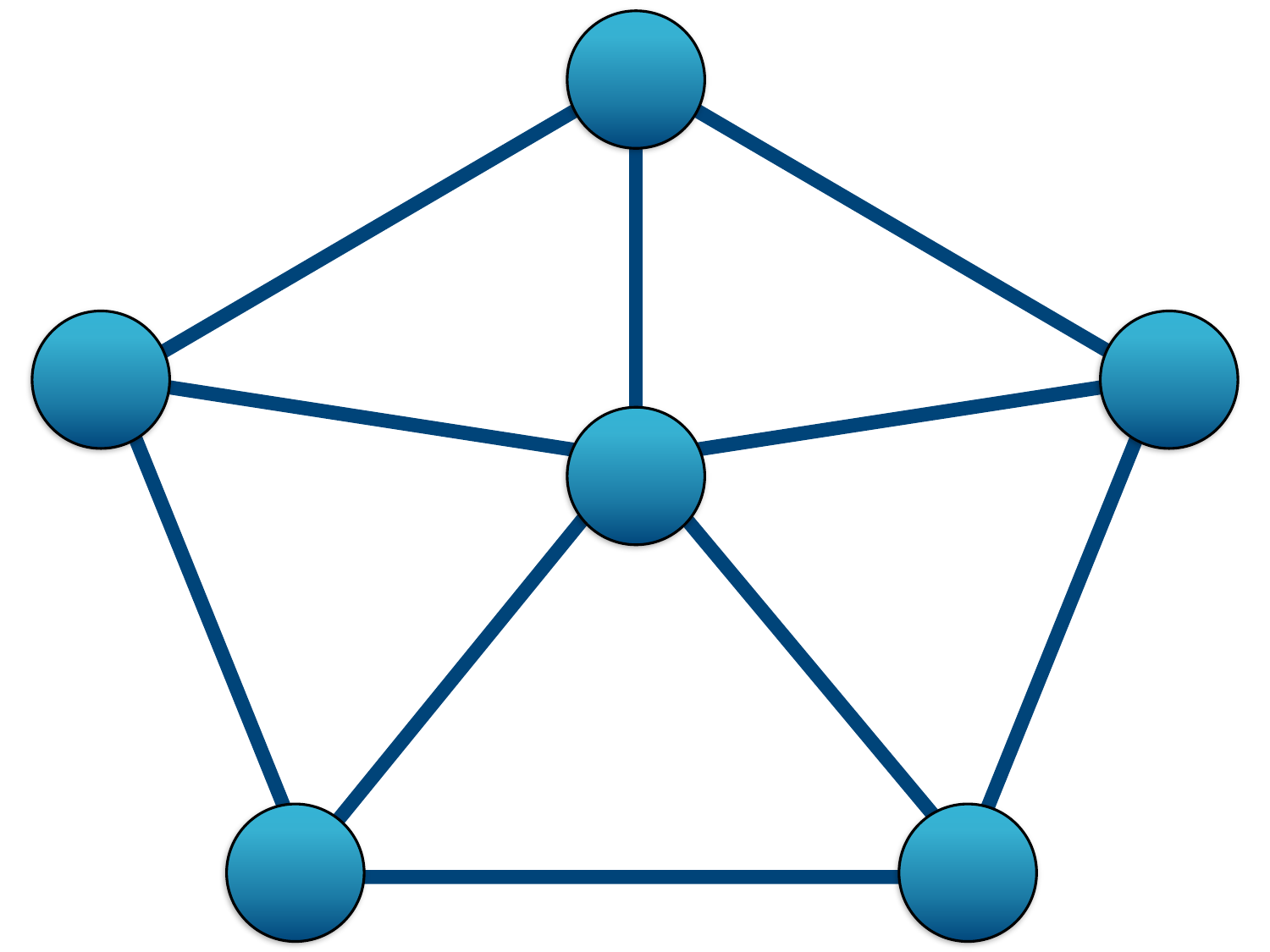}} & 
 $n =6$ &  
 $\dloc \geq 2$ &
2 & 
 yes  
      \\ 
       \hline
             \parbox[c]{16em}{\includegraphics[width=0.85in]{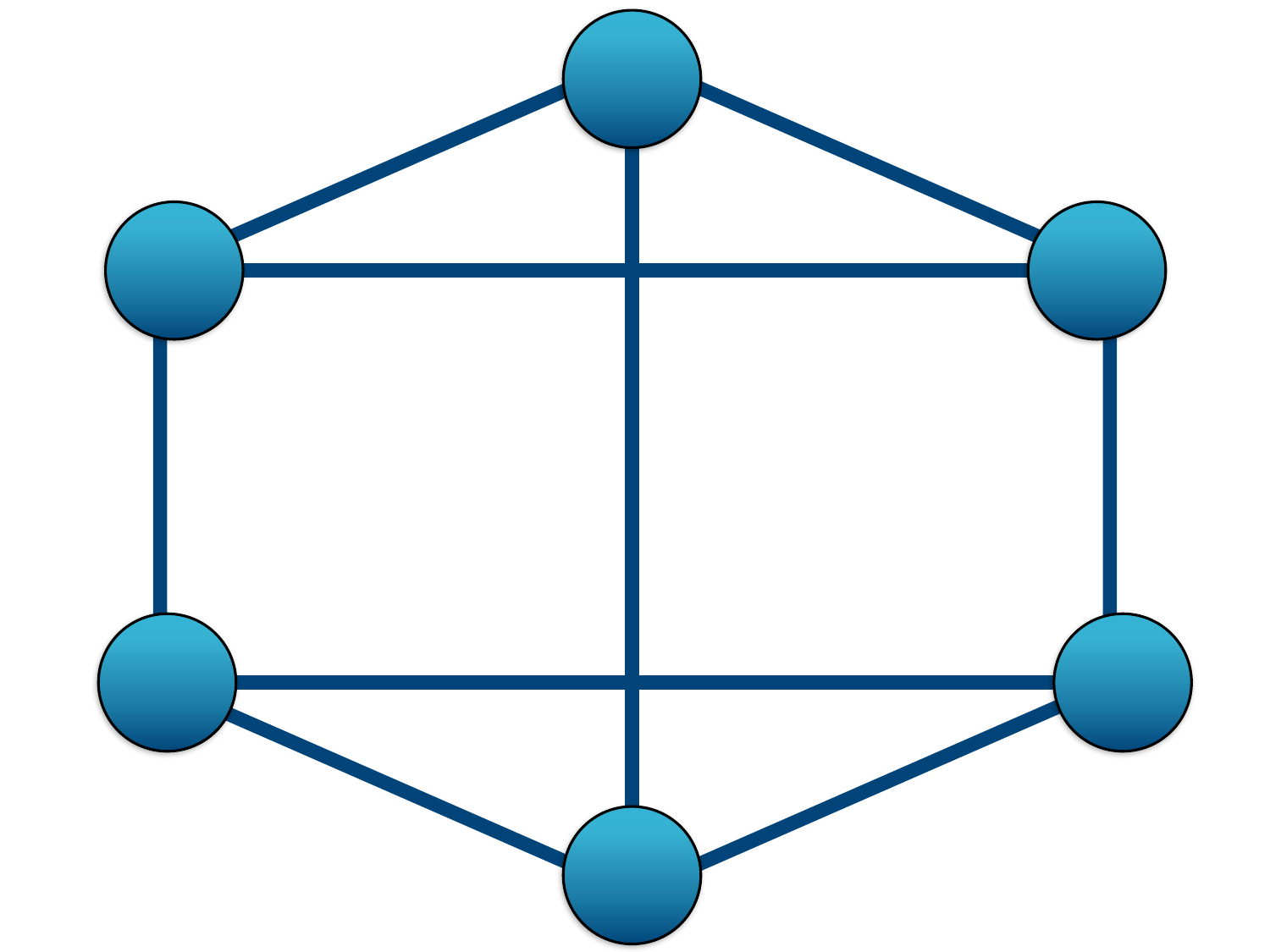}} & 
 $n =6$ &  
 $\dloc \geq 2$ &
 3 & 
 no 
 \\
        \hline
             \parbox[c]{16em}{\includegraphics[width=0.9in]{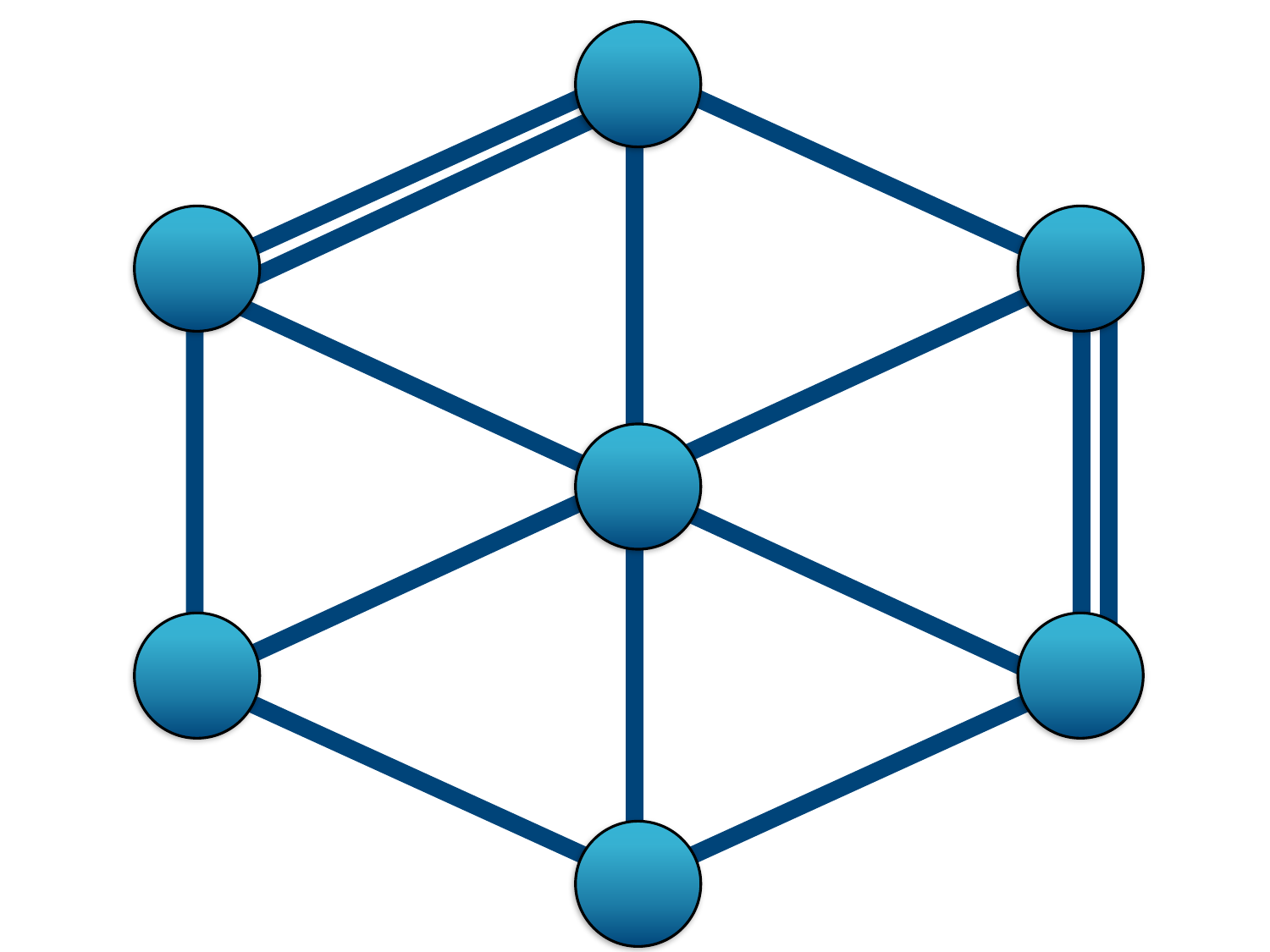}} & 
 $n =7$ &  
 $\dloc =3$ &
 2 & 
 yes  
 \\
 \hline
\end{tabular}
\end{center}
 \caption{\label{tabal-summary} 
 Summary of the generation protocols of various multi-photon qudit graph states presented in this work. For each protocol, the corresponding graph representation of the target state is shown, along with the number of photons it contains, the number $q$ of photonic time-bin states, the number of quantum emitters (with $q$ ground states each) needed to produce the state, and whether or not photon-emitter interaction is required for the protocol.}
\end{table}
\end{widetext}
 
A summary of the multi-qudit graph state generation protocols presented so far is given in Table.~\ref{tabal-summary}.

\section{Quantum error correcting codes}\label{sec:QECCs}

It is known that AME states are useful for constructing QECCs \cite{Scott,Zahra-minimalsupport,Zahra-MShortening}. In this section, we show how this connection can be exploited to develop protocols for generating multiphoton logical states of QECCs. First, we briefly review the relation between certain QECCs and AME states.
A subspace $\mathcal{C}$ spanned by an orthonormal set of states $\{\ket{\psi_0}, \ket{\psi_1}, \dots , \ket{\psi_{\dloc^k -1}}\}$, also called \emph{codewords}, is a QECC with parameters $[\![n,k,d]\!]_\dloc$.
This code is a $q^k$ dimensional subspace that encodes $k$ logical qudits into $n$ physical qudits, if it obeys the Knill-Laflamme conditions \cite{Knill,Gottesman-thesis}
\begin{equation}
  \forall m,m' \in [\dloc^k] \colon\ \bra{\psi_m}E^\dagger F\ket{\psi_{m'}}=f(E^{\dagger}F)\,\delta_{m,m'},
\end{equation}
for all errors $E,F$ with $\text{weight}(E^{\dagger}F) \leq d$, where the weight of an operator is defined to be the number of sites on which it acts non-trivially.
The parameter $d$ is the distance of the quantum code, which is the minimal number of single-qudit operations that are needed to
create a non-zero overlap between any two codewords from the code state space $\mathcal{C}$.
A QECC with minimum distance $d$ can correct errors that affect no more than $(d-1)/2$ of the physical qudits.
In this section, we show how to generate a $[\![3,1,2]\!]_3$ QECC, for which the codewords are all AME states of $3$ qutrits. 

First, let us review how to construct the $[\![3,1,2]\!]_3$ quantum code  (see also \cite{Calderbank,Ketkar2005,Zahra-MShortening}).
It is known that the $\AME(3,3)$ states  \cite{Alsina-QECCs,Zahra-MShortening}
{\begin{equation}\label{eq:codewords[[3,2,1]]}
   \begin{split}
    \ket{\psi_0} &= \sum_{j=0}^2 \ket{j, j, j}, \\
    \ket{\psi_1} &= M \ket{\psi_0} = \sum_{j=0}^2 \ket{j+1\ \mathrm{mod}\ 3, j, j+2\ \mathrm{mod}\ 3}, \\
    \ket{\psi_2} &= M^2 \ket{\psi_0} = \sum_{j=0}^2 \ket{j+2\ \mathrm{mod}\ 3, j, j+1\ \mathrm{mod}\ 3},
   \end{split}
 \end{equation}}
are the codewords of the $[\![3,1,2]\!]_3$ code.
In the above, the operator $M$ is defined as $M= X \otimes \1 \otimes X^2$.

In order to construct the corresponding graph states, it helps to first notice that by performing an $H$ gate, Eq.~\eqref{eq:H-on-qudits}, on the first and third qudits, we get
 \begin{equation}
 \ket{\psi'_0} = H \otimes \1 \otimes H \, \ket{\psi_0} = \sum_{i,j,k=0}^2 \omega^{kj} \omega^{ij} \ \ket{k,j,i},
 \end{equation}
 \begin{equation}
   \ket{\psi'_1} = H \otimes \1 \otimes H \, \ket{\psi_1} = \sum_{i,j,k=0}^2 \omega^{(j+1)k} \omega^{(j+2)i}\ \ket{k,j,i},
   \end{equation}
   \begin{equation}
   \ket{\psi'_2} = H \otimes \1 \otimes H \, \ket{\psi_2} = \sum_{i,j,k=0}^2 \omega^{(j+2)k} \omega^{(j+1)i} \ \ket{k,j,i} \, .
 \end{equation}
Notice that the state $\ket{\psi'_0}$ is equivalent to the 3-qutrit graph state shown in Fig.~\ref{fig:qutrit-cluster}(a) and discussed in Sec.~\ref{sec:Qutrit-clusterstates}. Thus, we already know how to generate this state, and so it remains 
to show how to produce $\ket{\psi'_1}$ and $\ket{\psi'_2}$ from a 3-level quantum emitter.
For this, let us first discuss how to generate $\ket{\psi'_1}$: 
{
 \begin{itemize}[leftmargin=.5in]
   \item[Step 1.] Prepare the emitter in the state\\ 
   $\hookrightarrow  \ket{\phi\step1} =\ket{2}_\qd$
   
  \item[Step 2.] $H$ gate, Eq.~\eqref{eq:H-on-qudits}, on the emitter\\
   $\hookrightarrow  \ket{\phi\step2} =\sum_{i=0}^2\omega^{2i} \ket{i}_{\qd}$
   
  \item[Step 3.] ${\cal P}_\mathrm{pump}$, Eq.~\eqref{eq:photon-pumping}, on the emitter\\
  $\hookrightarrow  \ket{\phi\step3} =\sum_{i=0}^2\omega^{2i} \ket{i}_\qd\ket{i}_{\p_1}$

  \item[Step 4.] $H$ gate on the emitter\\
   $   \hookrightarrow  \ket{\phi\step4} = \sum_{i,j=0}^2 \omega^{2i} \omega^{ij} \ket{j}_\qd\ket{i}_{\p_1}$

    \item[Step 5.] ${\cal P}_\mathrm{pump}$ and $H$ gate on the emitter\\
      $   \hookrightarrow  \ket{\phi\step5} = \sum_{i,j,k=0}^2 \omega^{2i} \omega^{ij} \omega^{jk} \ket{k}_\qd \ket{j}_{\p_2} \ket{i}_{\p_1}$
  
    \item[Step 6.] $Z$ gate, Eq.~\eqref{eq:defZ}, on the emitter\\
    $   \hookrightarrow  \ket{\phi\step6} =\\\sum_{i,j,k=0}^2 \omega^{2i} \omega^{ij} \omega^{jk} \omega^{k} \, \ket{k}_{\qd}\ket{j}_{\p_2} \ket{i}_{\p_1}$
    
    \item[Step 7.] ${\cal P}_\mathrm{pump}$ and $H$ gate on the emitter\\
    $  \hookrightarrow  \ket{\phi\step7} =$\\
    $\sum_{\substack{i,j,\\k,l=0}}^2 \omega^{2i} \omega^{ij} \omega^{jk} \omega^{k} \omega^{kl}\, \ket{l}_{\qd} \ket{k}_{\p_3}\ket{j}_{\p_2} \ket{i}_{\p_1}$
    
  \item[Step 8.] Measure the emitter in the $Z$ basis. Perform the gate $Z_{\p_3}^{2o}$ on photon $\p_3$, where $o$ is the measurement outcome. The resulting state is the desired $\ket{\psi'_1}$:
  \begin{align}
  \begin{split}
     \hookrightarrow  \ket{\phi\step8} &=\ket{\psi'_1} \\
     &= \sum_{i,j,k=0}^2 \omega^{2i} \omega^{ij}  \omega^{jk} \omega^{k} \, \ket{k}_{\p_3}\ket{j}_{\p_2} \ket{i}_{\p_1}.
     \end{split}
  \end{align}
\end{itemize}
}

A similar protocol can be used to generate $\ket{\psi'_2}$:
{
 \begin{itemize}[leftmargin=.5in]
   \item[Step 1.] Prepare the emitter in the state\\
    $\hookrightarrow  \ket{\phi\step1} =\ket{1}_\qd$
    
  \item[Step 2.]  $H$ gate, Eq.~\eqref{eq:H-on-qudits}, on the emitter\\
  $\hookrightarrow  \ket{\phi\step2} = \sum_{i=0}^2\omega^{i} \ket{i}_{\qd} $
  
  \item[Step 3.]  ${\cal P}_\mathrm{pump}$, Eq.~\eqref{eq:photon-pumping}, to obtain the state\\
  $ \hookrightarrow  \ket{\phi\step3} =\sum_{i=0}^2\omega^{i} \ket{i}_\qd\ket{i}_{\p_1}$

  \item[Step 4.] $H$ gate on the emitter\\
  $\hookrightarrow  \ket{\phi\step4} =\sum_{i,j=0}^2 \omega^{i} \omega^{ij} \ket{j}_\qd\ket{i}_{\p_1}$
  
   \item[Step 5.]   ${\cal P}_\mathrm{pump}$ and $H$ gate on the emitter\\
   $\hookrightarrow  \ket{\phi\step5} =\sum_{i,j,k=0}^2 \omega^{i} \omega^{ij} \omega^{jk} \, \ket{k}_{\qd}\ket{j}_{\p_2} \ket{i}_{\p_1}$
   
    \item[Step 6.] $Z^2$ gate, Eq.~\eqref{eq:defZqudit}, on the emitter\\
     $\hookrightarrow  \ket{\phi\step6} =\sum_{i,j,k=0}^2 \omega^{i} \omega^{ij} \omega^{jk} \omega^{2k} \, \ket{k}_{\qd}\ket{j}_{\p_2} \ket{i}_{\p_1}$
      
   \item[Step 7.]  ${\cal P}_\mathrm{pump}$ and $H$ gate on the emitter\\
    $\hookrightarrow  \ket{\phi\step7} =$\\$\sum_{\substack{i,j,\\k,l=0}}^2 \omega^{i} \omega^{ij} \omega^{jk} \omega^{2k} \omega^{kl}\, \ket{l}_{\qd} \ket{k}_{\p_3}\ket{j}_{\p_2} \ket{i}_{\p_1}$
    
  \item[Step 8.] Measure the emitter in the $Z$ basis. Perform the gate $Z_{\p_3}^{2o}$ on photon $\p_3$, where $o$ is the measurement outcome. The resulting state us the desired $\ket{\psi'_2}$:
  \begin{align}
  \begin{split}
   \hookrightarrow  \ket{\phi\step8} &=\ket{\psi'_2} \\
   &= \sum_{i,j,k=0}^2 \omega^{i} \omega^{ij}  \omega^{jk} \omega^{2k} \, \ket{k}_{\p_3}\ket{j}_{\p_2} \ket{i}_{\p_1}\ .
  \end{split}
  \end{align}
\end{itemize}
}

Similar protocols can be devised to generate the codewords of QECCs associated with any of the AME states discussed in Sec.~\ref{sec:AME-states}.

\section{Physical implementations}\label{sec:implementations}

Photonic qudit states can be generated from a range of different emitters. On the solid-state side, an example is the well-known NV center in diamond, where among its three ground states, $\ket{0}$ and $\ket{\pm1}$, state $\ket{0}$ can be reliably pumped to an excited state that subsequently decays back down to $\ket{0}$, emitting a photon \cite{DOHERTY20131,Wolfowicz2021}. NV centers can thus be used to generate photonic time-bin qutrits. Since however the NV has a very low probability of emitting into the zero-phonon line, it may be preferable to consider alternative defects. The silicon-carbon divacancy in SiC is another defect that has a triplet ground state, and that generally resembles the electronic structure of NV-diamond, but with improved optical properties and comparably long coherence times \cite{Christle2015,Nagy2018,Nagy2019}. For local dimension $q=4$, the silicon vacancy in SiC could be used instead.   

Atomic systems provide even more options for the value of $q$, as they can contain well-resolved hyperfine states. Trapped ions such as $\ce{^{40}Ca+}$ \cite{Ringbauer-ION-2021} and $\ce{^{171}Yb+}$ \cite{Timoney-171Yb, Webster-171Yb,Aksenov-ION-2022} can serve as multi-level quantum emitters with local dimensions ranging from $q=3$ up to $q=7$~\cite{Timoney-171Yb, Webster-171Yb,Randall-PRA-Ion,Senko-PRX-Ion,Ringbauer-ION-2021,Aksenov-ION-2022}. There has been significant experimental progress in realizing single- and multi-qudit gate operations in chains of such ions~\cite{Ringbauer-ION-2021,Aksenov-ION-2022}. Few-atom systems in a cavity are another contender for generation of qudit graph states. A recent milestone experiment demonstrated the generation of GHZ and linear cluster states of qubits from a Rb atom~\cite{Rempe-Nature2022}. 
While in that experiment the photonic qubit was encoded in the polarization degree of freedom, and therefore only two of the states in the ground state manifold were used explicitly in the protocol, a similar setup could be employed to demonstrate qudit graph-state generation with time-bin encoding and with more levels participating in the generation process (up to $q=8$ when both the $F=1$ and the $F=2$ manifolds are used). To create more complex qudit graphs, cavity-mediated interactions between two or more atoms can be leveraged to create qudit $CZ$-type gates  by modifying the protocol for the already demonstrated two-qubit gates in such systems~\cite{Welte-PRX2018}.

\section{Conclusions and outlook}\label{sec:conclusion}
In conclusion, we presented explicit protocols for using coupled, controllable quantum emitters to deterministically generate multi-photon entangled states of time-bin qudits. {We proved that any graph state of time-bin photonic qudits can be generated from coupled quantum emitters with a suitable level structure using a small set of single- and two-qubit gates on the emitters. We showed that any linear graph state can be produced from a single quantum emitter, and that 2D cluster states of size $m\times n$ can be generated using $m<n$ emitters.}
We then focused on the problem of generating highly entangled states known as AME states, which are important for a number applications in quantum networks and quantum error correction. We showed that such states can be produced from a small number of emitters, with or without photon-emitter interaction. In some cases, we found that fewer emitters can be used if photon-emitter interaction is available, providing hints at what sort of resource tradeoffs are possible in general. Potential candidates include defect centers in solids as well as atomic systems. Our results provide a clear path forward toward the efficient generation of complex states of light for quantum information applications and a guide to experimental groups. 

Future directions this work opens include addressing the question of what the minimal resources are---in terms of the number of emitters and the circuit depth---for the generation of a target qudit graph. One could imagine developing a formalism to answer these questions along the lines of what was done in the context of qubits~\cite{Li2022}. Another interesting direction would be to design specific qudit gates for various candidate emitters and to quantify the anticipated performance of the protocols. Our protocols could also be combined with new approaches to photonic one-way quantum repeaters based on photonic qudit states~\cite{Zheng_2022} to build error-correction into that approach.

\acknowledgements
We would thank Karl Jansen, Chenxu Liu, and Evangelia Takou for discussions and useful comments. SEE acknowledges support by the ARO MURI Quantum Network Science: W91INF-21-2-0214.
EB acknowledges support by the National Science Foundation (grant nos. 1741656 and 2137953). This work was also supported in part by the Commonwealth Cyber Initiative, an investment in
the advancement of cyber R\&D, innovation, and workforce development.

\appendix

{
   
\section{How to generate ladder and  two-dimensional graph states} \label{app:ladder}

To generate a ladder graph state (see Fig.~\ref{fig:1D-2D}(a)), we start from two quantum emitters with $\dloc$ ground levels \cite{Sophia-2Dcluster}.
In this procedure, we only allow two-qubit gates that act on emitters (no emitter-photon interaction). 
The procedure is as follows: 
\begin{figure}
 \includegraphics[scale=0.22]{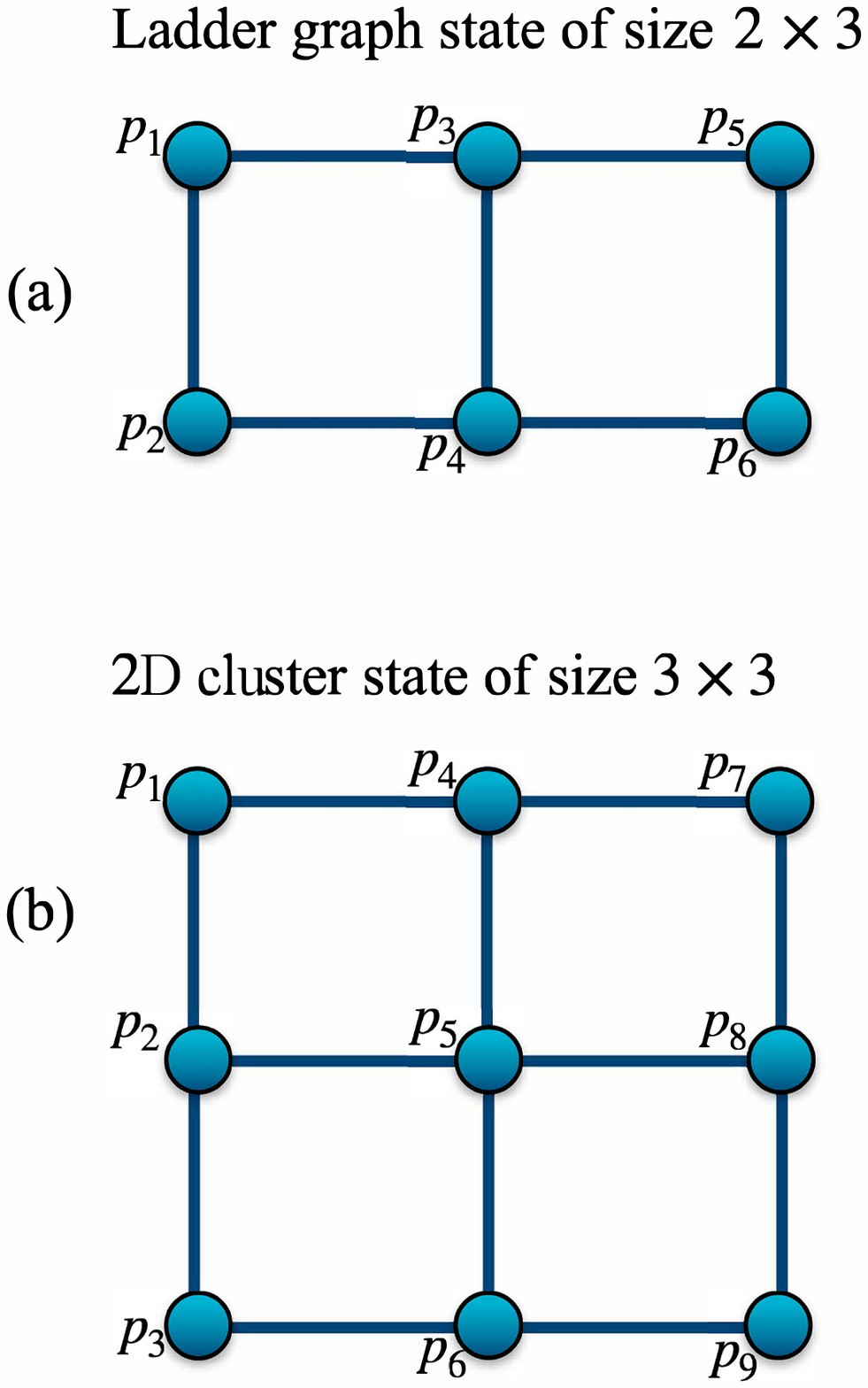}
 \centering
 \caption{\label{fig:1D-2D}  {
(a) Graph representing the ladder graph state of size $2 \times 3$, and (b) graph for the two-dimensional cluster state of size $3 \times 3$.}
}
\end{figure}

 \begin{itemize}[leftmargin=.5in]
  \item[Step 1.] Prepare two emitters in the state\\
   $\hookrightarrow \ket{\phi\step1}=   \ket{0}_{\qd_1} \ket{0}_{\qd_2}$
     \item[Step 2.] $H$ gate, Eq.~\eqref{eq:H-on-qudits}, on each emitter\\
   $\hookrightarrow  \ket{\phi\step2}= \sum_{i,j=0}^2 \ket{i}_{\qd_1} \ket{j}_{\qd_2}$
   
  \item[Step 3.] $CZ$ gate, Eq.~\eqref{eq:CZqudit}, between the emitters\\
   $\hookrightarrow  \ket{\phi\step3}= \sum_{i,j=0}^{\dloc-1} \omega^{ij}\ \ket{i}_{\qd_1} \ket{j}_{\qd_2}$
   
    \item[Step 4.] ${\cal P}_\mathrm{pump}$, Eq.~\eqref{eq:photon-pumping}, on each emitter\\
   $\hookrightarrow  \ket{\phi\step4}= \sum_{i,j=0}^{\dloc-1} \omega^{ij}\ \ket{i}_{\qd_1} \ket{i}_{\p_1} \ket{j}_{\qd_2} \ket{j}_{\p_2}$
   
    \item[Step 5.] $H$ gate on each emitter\\
   $\hookrightarrow  \ket{\phi\step5}= \\ \sum_{\substack{i,j, \\ k,l=0}}^{\dloc-1} \omega^{ij} \omega^{ik} \omega^{jl}\ \ket{k}_{\qd_1} \ket{i}_{\p_1} \ket{l}_{\qd_2} \ket{j}_{\p_2}$
   
    \item[Step 6.] ${\cal P}_\mathrm{pump}$ on each emitter\\
   $\hookrightarrow  \ket{\phi\step6}=$\\$ \sum_{\substack{i,j, \\ k,l=0}}^{\dloc-1}  \omega^{ij} \omega^{ik} \omega^{jl}\ \ket{k}_{\qd_1} \ket{k}_{\p_3} \ket{i}_{\p_1} \ket{l}_{\qd_2} \ket{l}_{\p_4} \ket{j}_{\p_2}$
   
      \item[Step 7.] $CZ$ between emitters\\
   $\hookrightarrow  \ket{\phi\step6}=$ \\$ \sum_{\substack{i,j, \\ k,l=0}}^{\dloc-1} \omega^{ij} \omega^{ik} \omega^{jl} \omega^{kl}\ket{k}_{\qd_1} \ket{k}_{\p_3} \ket{i}_{\p_1} \ket{l}_{\qd_2} \ket{l}_{\p_4} \ket{j}_{\p_2}$

    \item[Step 8.] $H$ gate on each emitter\\
   $\hookrightarrow  \ket{\phi\step8} $\\
   $= \sum_{\substack{i,j,k \\ l,m,n=0}}^{\dloc-1} \kappa \ \ket{m}_{\qd_1} \ket{k}_{\p_3} \ket{i}_{\p_1} \ket{n}_{\qd_2} \ket{l}_{\p_4} \ket{j}_{\p_2}$\\
   where $\kappa \coloneqq \omega^{ij} \omega^{ik} \omega^{jl} \omega^{kl} \omega^{km} \omega^{ln}$

     \item[Step 9.] $CZ$ gate between emitters\\
   $\hookrightarrow  \ket{\phi\step9}=$
   \\
   $ \sum_{\substack{i,j,k \\ l,m,n=0}}^{\dloc-1} \kappa'\ \ket{m}_{\qd_1} \ket{k}_{\p_3} \ket{i}_{\p_1} \ket{n}_{\qd_2} \ket{l}_{\p_4} \ket{j}_{\p_2}$\\
   where $\kappa' \coloneqq \omega^{ij} \omega^{ik} \omega^{jl} \omega^{kl}\omega^{km} \omega^{ln} \omega^{mn}$

       \item[Step 10.] ${\cal P}_\mathrm{pump}$ and then $H$ gate on each emitter\\
 \begin{align*}
 \begin{split}
& \hookrightarrow  \ket{\phi\step{10}} =\\
&\sum_{\substack{i,j,k,l \\ m,n,o,r=0}}^{\dloc-1} \kappa''\ \ket{o}_{\qd_1} \ket{m}_{\p_5} \ket{k}_{\p_3} \ket{i}_{\p_1} \ket{r}_{\qd_2} \ket{n}_{\p_6} \ket{l}_{\p_4} \ket{j}_{\p_2}
 \end{split}
\end{align*}        
  where $\kappa'' \coloneqq \omega^{ij} \omega^{ik} \omega^{jl} \omega^{kl} \omega^{km} \omega^{ln} \omega^{mn} \omega^{mo} \omega^{nr}$ 
  
 \item[Step 11.]  Measure each emitter in the $Z$ basis. \\
 Perform the local gates $Z_{\p_5}^{(\dloc-1)o_1}Z_{\p_6}^{(\dloc-1)o_2}$ on photons 5 and 6, where $o_1$ and $o_2$ are the measurement outcomes for emitters 1 and 2, respectively, yielding  
     \begin{align*}
 \begin{split}
& \hookrightarrow  \ket{\phi\step{11}} \\
&=\sum_{\substack{i,j,k \\ l,m,n=0}}^{\dloc-1}  \kappa''' \  \ket{m}_{\p_5} \ket{k}_{\p_3} \ket{i}_{\p_1} \ket{n}_{\p_6} \ket{l}_{\p_4} \ket{j}_{\p_2},
 \end{split}
\end{align*}        
 \end{itemize}
 where $\kappa''' \coloneqq  \omega^{ij} \omega^{ik} \omega^{jl} \omega^{kl} \omega^{km} \omega^{ln} \omega^{mn}$.
  
The final state is a qudit photonic ladder graph state like that shown in Fig.~\ref{fig:1D-2D}(a).
Note that to produce similar ladder graph states of larger size, i.e, size $2 \times n$, we need to repeat Steps 6-8.

To extend a ladder and construct a two-dimensional graph state of size $m \times n$ (also known as a cluster state), we need $m$ quantum emitters. As an example, we list the steps one needs to follow to generate a $3 \times 3$ graph state (see Fig.~\ref{fig:1D-2D}(b)) using three quantum emitters:
 \begin{itemize}[leftmargin=.5in]
    \begin{widetext}
  \item[Step 1.] Prepare three emitters in the state\\
   $\hookrightarrow \ket{\phi\step1}=   \ket{0}_{\qd_1} \ket{0}_{\qd_2} \ket{0}_{\qd_3}$
     \item[Step 2.] $H$ gate, Eq.~\eqref{eq:H-on-qudits}, on each emitter\\
   $\hookrightarrow  \ket{\phi\step2}= \sum_{i,j,k=0}^2 \ket{i}_{\qd_1} \ket{j}_{\qd_2} \ket{k}_{\qd_3}$
   
  \item[Step 3.] Two $CZ$ gates, Eq.~\eqref{eq:CZqudit}, between $\qd_1 \qd_2$ and $\qd_2 \qd_3$\\
   $\hookrightarrow  \ket{\phi\step3}= \sum_{i,j,k=0}^{\dloc-1} \omega^{ij} \omega^{jk}\ \ket{i}_{\qd_1} \ket{j}_{\qd_2} \ket{k}_{\qd_3}$
  
    \item[Step 4.] ${\cal P}_\mathrm{pump}$, Eq.~\eqref{eq:photon-pumping}, on each emitter\\
   $\hookrightarrow  \ket{\phi\step4}= \sum_{i,j,k=0}^{\dloc-1} \omega^{ij} \omega^{jk}\ \ket{i}_{\qd_1} \ket{i}_{\p_1} \ket{j}_{\qd_2} \ket{j}_{\p_2}  \ket{k}_{\qd_3} \ket{k}_{\p_3} $
   
    \item[Step 5.] $H$ gate on each emitter\\
   $\hookrightarrow  \ket{\phi\step5}= \sum_{\substack{i,j,k, \\ l,m,n=0}}^{\dloc-1} \omega^{ij} \omega^{jk} \omega^{il} \omega^{jm} \omega^{kn}\ \ket{l}_{\qd_1} \ket{i}_{\p_1} \ket{m}_{\qd_2} \ket{j}_{\p_2}  \ket{n}_{\qd_3} \ket{k}_{\p_3} $
   
    \item[Step 6.] Two $CZ$ gates, Eq.~\eqref{eq:CZqudit}, between $\qd_1 \qd_2$ and $\qd_2 \qd_3$\\
   $\hookrightarrow  \ket{\phi\step6}= \sum_{\substack{i,j,k, \\ l,m,n=0}}^{\dloc-1} \omega^{ij} \omega^{jk} \omega^{il} \omega^{jm} \omega^{kn} \omega^{lm} \omega^{mn} \ \ket{l}_{\qd_1} \ket{i}_{\p_1} \ket{m}_{\qd_2} \ket{j}_{\p_2}  \ket{n}_{\qd_3} \ket{k}_{\p_3} $
   
      \item[Step 7.] ${\cal P}_\mathrm{pump}$ on each emitter\\
  $\hookrightarrow  \ket{\phi\step7}= \sum_{\substack{i,j,k, \\ l,m,n=0}}^{\dloc-1} \omega^{ij} \omega^{jk} \omega^{il} \omega^{jm} \omega^{kn} \omega^{lm} \omega^{mn} \ \ket{l}_{\qd_1} \ket{l}_{\p_4} \ket{i}_{\p_1} \ket{m}_{\qd_2} \ket{m}_{\p_5} \ket{j}_{\p_2}  \ket{n}_{\qd_3} \ket{n}_{\p_6}\ket{k}_{\p_3} $

    \item[Step 8.] $H$ gate on each emitter\\
   $\hookrightarrow  \ket{\phi\step8}= \sum_{\substack{i,j,k \\ l,m,n, \\o,r,s=0}}^{\dloc-1} \omega^{ij} \omega^{jk} \omega^{il} \omega^{jm} \omega^{kn} \omega^{lm} \omega^{mn} \omega^{lo} \omega^{mr} \omega^{ns}  \ \ket{o}_{\qd_1} \ket{l}_{\p_4} \ket{i}_{\p_1} \ket{r}_{\qd_2} \ket{m}_{\p_5} \ket{j}_{\p_2}  \ket{s}_{\qd_3} \ket{n}_{\p_6}\ket{k}_{\p_3}$

     \item[Step 9.]  Two $CZ$ gates, Eq.~\eqref{eq:CZqudit}, between $\qd_1 \qd_2$ and $\qd_2 \qd_3$\\
   $\hookrightarrow  \ket{\phi\step9}= \sum_{\substack{i,j,k \\ l,m,n, \\o,r,s=0}}^{\dloc-1} \omega^{ij} \omega^{jk} \omega^{il} \omega^{jm} \omega^{kn} \omega^{lm} \omega^{mn} \omega^{lo} \omega^{mr} \omega^{ns}  \omega^{or} \omega^{rs} \ \ket{o}_{\qd_1} \ket{l}_{\p_4} \ket{i}_{\p_1} \ket{r}_{\qd_2} \ket{m}_{\p_5} \ket{j}_{\p_2}  \ket{s}_{\qd_3} \ket{n}_{\p_6}\ket{k}_{\p_3}$

       \item[Step 10.] ${\cal P}_\mathrm{pump}$ on each emitter\\
   $\hookrightarrow  \ket{\phi\step{10}}= \sum_{\substack{i,j,k \\ l,m,n, \\o,r,s=0}}^{\dloc-1} \xi \ \ket{o}_{\qd_1}  \ket{ o}_{\p_7} \ket{l}_{\p_4} \ket{i}_{\p_1} \ket{r}_{\qd_2} \ket{ r}_{\p_8} \ket{m}_{\p_5} \ket{j}_{\p_2}  \ket{s}_{\qd_3} \ket{s }_{\p_9} \ket{n}_{\p_6}\ket{k}_{\p_3}$
  
  where $\xi \coloneqq \omega^{ij} \omega^{jk} \omega^{il} \omega^{jm} \omega^{kn} \omega^{lm} \omega^{mn} \omega^{lo} \omega^{mr} \omega^{ns}  \omega^{or} \omega^{rs} $

 \item[Step 11.] $H$ gate on each emitter\\
   $\hookrightarrow  \ket{\phi\step{11}}= \sum_{\substack{i,j,k, l,m,n, \\o,r,s,t,x,y=0}}^{\dloc-1} \xi' \ \ket{t}_{\qd_1}  \ket{ o}_{\p_7} \ket{l}_{\p_4} \ket{i}_{\p_1} \ket{x}_{\qd_2} \ket{ r}_{\p_8} \ket{m}_{\p_5} \ket{j}_{\p_2}  \ket{y}_{\qd_3} \ket{s }_{\p_9} \ket{n}_{\p_6}\ket{k}_{\p_3}$
  
  where $\xi' \coloneqq \omega^{ij} \omega^{jk} \omega^{il} \omega^{jm} \omega^{kn} \omega^{lm} \omega^{mn} \omega^{lo} \omega^{mr} \omega^{ns}  \omega^{or} \omega^{rs} \omega^{ot} \omega^{rx} \omega^{sy} $ 
  
 \item[Step 12.]  Measure each emitter in the $Z$ basis. \\
 Perform the local gates $Z_{\p_7}^{(\dloc-1)o_1}Z_{\p_8}^{(\dloc-1)o_2} Z_{\p_9}^{(\dloc-1)o_3}$ on photons 7, 8 and 9, where $o_1$, $o_2$ and $o_3$ are the measurement outcomes for emitters 1, 2 and 3, respectively, yielding  
     \begin{align*}
 \begin{split}
& \hookrightarrow  \ket{\phi\step{12}} \\
&=\sum_{\substack{i,j,k,l, \\ m,n,o,r,s=0}}^{\dloc-1}  \xi'' \   \ket{ o}_{\p_7} \ket{l}_{\p_4} \ket{i}_{\p_1} \ket{ r}_{\p_8} \ket{m}_{\p_5} \ket{j}_{\p_2}  \ket{s }_{\p_9} \ket{n}_{\p_6}\ket{k}_{\p_3},
 \end{split}
\end{align*}  
 where $\xi'' \coloneqq   \omega^{ij} \omega^{jk} \omega^{il} \omega^{jm} \omega^{kn} \omega^{lm} \omega^{mn} \omega^{lo} \omega^{mr} \omega^{ns}  \omega^{or} \omega^{rs}$.
 \end{widetext}      
 \end{itemize}

And, to produce similar graph states of larger size, i.e., size $3 \times n$, we need to repeat Steps 9-11.

\section{How to generate AME states of $5$ qudits}\label{app:AME5q}

Here, we discuss how to generate $\ket{\AME(5,\dloc)_{\text{graph}}}$ states. The graph and generation circuits are shown in Fig.~\ref{fig:AME5,q}.
We begin by presenting the protocol that requires only one quantum emitter with $\dloc$ ground levels (Fig.~\ref{fig:AME5,q}(b)): 
 \begin{itemize}[leftmargin=.5in]
  \item[Step 1.] Prepare the emitter in the state \\
  $\hookrightarrow \ket{\phi\step1} =\ket{0}_\qd$
  
    \item[Step 2.] Perform a Hadamard gate $H$, Eq.~\eqref{eq:H-on-qudits}, to produce the state\\
     $ \hookrightarrow \ket{\phi\step2} = \ket{X_0} = \sum_{i} \ket{i}_{\qd}$
     
   \item[Step 3.] Perform a photon-pumping operation ${\cal P}_\mathrm{pump}$, Eq.~\eqref{eq:photon-pumping}, to generate a time-bin encoded photon, yielding the state \\$ \hookrightarrow \ket{\phi\step3} =\sum_{i=0}^{\dloc -1} \ket{i}_{\qd} \ket{i}_{\p_1} $ 
   
    \item[Step 4.] Perform an $H$ gate on the emitter to get\\
     $ \hookrightarrow \ket{\phi\step4} =\sum_{i,j=0}^{\dloc -1} \omega^{ij} \, \ket{j}_{\qd} \ket{i}_{\p_1} $
     
    \item[Step 5.] Repeat steps 3 and 4 three more times to obtain\\ $\hookrightarrow \ket{\phi\step5} =\sum_{i,j,k,l,m=0}^{\dloc -1}$\\ $\omega^{ij} \omega^{jk} \omega^{kl} \omega^{lm} \ \ket{m}_{\qd} \ket{l}_{\p_4} \ket{k}_{\p_3} \ket{j}_{\p_2} \ket{i}_{\p_1} $
    
     \item[Step 6.] Interact the first photon $\p_1$ with the emitter to perform a $CZ$ gate between them, yielding\\ 
     $\hookrightarrow \ket{\phi\step6} =\\ \sum_{i,j,k,l,m} \omega^{ij} \omega^{jk} \omega^{kl} \omega^{lm}  \omega^{mi}  \ket{m}_{\qd} \ket{l}_{\p_4} \ket{k}_{\p_3} \ket{j}_{\p_2} \ket{i}_{\p_1} 
    $
    
     \item[Step 7.] Perform a photon-pumping operation ${\cal P}_\mathrm{pump}$ followed by an $H$ gate on the emitter to obtain \\ 
     $\hookrightarrow \ket{\phi\step7} =$
     \[\sum_{i,j,k,l,m,r} \omega^{ij} \omega^{jk} \omega^{kl} \omega^{lm}  \omega^{mi} \omega^{mr}  \ket{r}_{\qd} \ket{m}_{\p_5} \ket{l}_{\p_4} \ket{k}_{\p_3} \ket{j}_{\p_2} \ket{i}_{\p_1}\]
     
     \item[Step 8.] Measure the emitter in the $Z$ basis and perform $Z_{\p_5}^{(q-1)o}$ on photon $\p_5$, where $o$ is the measurement outcome. The resulting state is the desired AME state:
     \begin{equation}
   \begin{split}
 &\hookrightarrow \ket{\phi\step8} =\ket{\AME(5,\dloc)} =\\
 & \sum_{i,j,k,l,m} \omega^{ij} \omega^{jk} \omega^{kl} \omega^{lm}  \omega^{mi}  \ket{m}_{\p_5} \ket{l}_{\p_4} \ket{k}_{\p_3} \ket{j}_{\p_2} \ket{i}_{\p_1} .\nonumber
    \end{split}
     \end{equation}
 \end{itemize}
 
In the protocol described above and shown in Fig.~\ref{fig:AME5,q}(b), we need to send one of the emitted photons back to interact with the emitter to generate a $CZ$ gate between them. 
An alternative approach that does not require emitter-photon interaction is possible if we have access to two coupled emitters with $\dloc$ ground levels each; one such protocol works as follows (see the quantum circuit in Fig.~\ref{fig:AME5,q}(c)):
 \begin{itemize}[leftmargin=.5in]
   \item[Step 1.] Prepare the two emitters in the state\\
    $\hookrightarrow \ket{\phi\step1} =\ket{0}_{\qd_1}\ket{0}_{\qd_2}$
    
   \item[Step 2.] Perform a Hadamard gate $H$, Eq.~\eqref{eq:H-on-qudits}, on each emitter, yielding\\
    $\hookrightarrow \ket{\phi\step2} =\sum_{i,j=0}^{\dloc -1} \ket{i}_{\qd_1}\ket{j}_{\qd_2}$
    
   \item[Step 3.] Perform a $CZ$ gate, Eq.~\eqref{eq:CZqudit}, on the two emitters to get\\ $\hookrightarrow \ket{\phi\step3} =\sum_{i,j=0}^{\dloc -1} \omega^{ij} \  \ket{i}_{\qd_1}\ket{j}_{\qd_2}$
   
   \item[Step 4.] Perform photon-pumping operations ${\cal P}_\mathrm{pump}$, Eq.~\eqref{eq:photon-pumping}, to each emitter to create two time-bin photonic qudits, yielding\\
   $\hookrightarrow \ket{\phi\step4} =\sum_{i,j=0}^{\dloc -1} \omega^{ij} \  \ket{i}_{\qd_1} \ket{i}_{\p_1} \ket{j}_{\qd_2} \ket{j}_{\p_2}$
   
   \item[Step 5.] Perform an $H$ gate on each emitter to get \\
   $ \hookrightarrow \ket{\phi\step5} = \\\sum_{i,j,k,l=0}^{\dloc -1} \omega^{ij}  \omega^{ik} \omega^{jl}\  \ket{k}_{\qd_1} \ket{i}_{\p_1} \ket{l}_{\qd_2} \ket{j}_{\p_2}$
   
   \item[Step 6.] Perform a photon-pumping operation on the first emitter ($\qd_1$) to produce another photon:\\
   $\hookrightarrow \ket{\phi\step6} =$\\
   $ \sum_{i,j,k,l=0}^{\dloc -1} \omega^{ij}  \omega^{ik} \omega^{jl}\  \ket{k}_{\qd_1} \ket{k}_{\p_3}\ket{i}_{\p_1} \ket{l}_{\qd_2} \ket{j}_{\p_2}$
   
   \item[Step 7.] Perform an $H$ gate on the first emitter ($\qd_1$) to get \\
   $\hookrightarrow \ket{\phi\step7} =$\\
   $ \sum_{i,j,k,l=0}^{\dloc -1} \omega^{ij}  \omega^{ik} \omega^{jl} \omega^{km}\  \ket{m}_{\qd_1} \ket{k}_{\p_3}\ket{i}_{\p_1} \ket{l}_{\qd_2} \ket{j}_{\p_2}$
   
   \item[Step 8.] Perform a $CZ$ gate on the two emitters to get\\
   $\hookrightarrow \ket{\phi\step8} =$
   \[\sum_{\substack{i,j,k, \\ l,m=0}}^{\dloc-1} \omega^{ij}  \omega^{ik} \omega^{jl} \omega^{km} \omega^{ml} \  \ket{m}_{\qd_1} \ket{k}_{\p_3} \ket{i}_{\p_1} \ket{l}_{\qd_2} \ket{j}_{\p_2} 
     \]
     
   \item[Step 9.] Perform a photon-pumping operation on each emitter, followed by an $H$ gate on each emitter:\\
   $\hookrightarrow \ket{\phi\step9} =$
   \[\sum_{\substack{i,j,k,l \\ m,r,s=0}}^{\dloc-1} \Omega \ket{r}_{\qd_1}\ket{m}_{\p_4} \ket{k}_{\p_3} \ket{i}_{\p_1} \ket{s}_{\qd_2}\ket{l}_{\p_5} \ket{j}_{\p_2},\] where $\Omega=\omega^{ij}  \omega^{ik} \omega^{jl} \omega^{km} \omega^{ml}\omega^{mr}\omega^{ls}$
   
   \item[Step 10.] Measure both emitters in the $Z$ basis and perform $Z_{\p_4}^{(q-1)o_1}Z_{\p_5}^{(q-1)o_2}$ on photons $\p_4$ and $\p_5$, where $o_1$ and $o_2$ are the measurement outcomes. This yields the desired 5-photon state:\\
   \begin{equation}
   \begin{split}
   &\hookrightarrow \ket{\phi\step{10}} =\ket{\AME(5,\dloc)_{\text{graph}}} = \\
   &\sum_{\substack{i,j,k, \\ l,m=0}}^{\dloc-1} \omega^{ij}  \omega^{ik} \omega^{jl} \omega^{km} \omega^{ml} \  \ket{m}_{\p_5} \ket{k}_{\p_3} \ket{i}_{\p_1} \ket{l}_{\p_4} \ket{j}_{\p_2} .\nonumber
      \end{split}
   \end{equation}
 \end{itemize}

 \section{How to generate AME states of $6$ qudits }\label{app:AME6q}
 
 Here, we discus the protocol for constructing the graph representation of the $\ket{\AME(6,\dloc)_{\text{graph}}}$ state shown in Fig.~\ref{fig:AME6,q}(a).
 This protocol requires two coupled emitters with $\dloc$ ground levels each. It also needs photon-emitter interaction.
This protocol consists of the following steps:
 \begin{itemize}[leftmargin=.5in]
   \item[Step 1.] Prepare two $\dloc$-level emitters in the state\\
    $\hookrightarrow \ket{\phi\step1} =\ket{0}_{\qd_1}\ket{0}_{\qd_2}$ 
    
   \item[Step 2.] Perform a Hadamard gate $H$, Eq.~\eqref{eq:H-on-qudits}, on each emitter to obtain\\
    $\hookrightarrow \ket{\phi\step2} =\sum_{i,j=0}^{\dloc -1} \ket{i}_{\qd_1}\ket{j}_{\qd_2}$
    
   \item[Step 3.] Perform a $CZ$ gate, Eq.~\eqref{eq:CZqudit}, on the two emitters to get\\ 
   $\hookrightarrow \ket{\phi\step3} =\sum_{i,j=0}^{\dloc -1} \omega^{ij} \  \ket{i}_{\qd_1}\ket{j}_{\qd_2}$
   
   \item[Step 4.] Perform photon-pumping operations ${\cal P}_\mathrm{pump}$, Eq.~\eqref{eq:photon-pumping}, on each emitter to produce two photons:\\
   $\hookrightarrow \ket{\phi\step4} =\sum_{i,j=0}^{\dloc -1} \omega^{ij} \  \ket{i}_{\qd_1} \ket{i}_{\p_1} \ket{j}_{\qd_2} \ket{j}_{\p_2}$
   
   \item[Step 5.] Perform an $H$ gate on each emitter to obtain \\
   $\hookrightarrow \ket{\phi\step5} = \\ \sum_{i,j,k,l=0}^{\dloc -1} \omega^{ij}  \omega^{ik} \omega^{jl}\  \ket{k}_{\qd_1} \ket{i}_{\p_1} \ket{l}_{\qd_2} \ket{j}_{\p_2}$
   
   \item[Step 6.] Apply two $CZ$ gates, one on the first photon ($\p_1$) and the second emitter ($\qd_2$), and the other on the two emitters, yielding \\
   $\hookrightarrow \ket{\phi\step6} = $\\
   $\sum_{i,j,k,l=0}^{\dloc -1} \omega^{ij}  \omega^{ik} \omega^{jl} \omega^{il} \omega^{kl}\  \ket{k}_{\qd_1} \ket{i}_{\p_1} \ket{l}_{\qd_2} \ket{j}_{\p_2}$
   
   \item[Step 7.] Perform photon-pumping operations on each emitter to create two more photons:\\ 
   $\hookrightarrow \ket{\phi\step7} = \sum_{i,j,k,l} $\\
   $ \omega^{ij}  \omega^{ik} \omega^{jl} \omega^{il} \omega^{kl}\ \ket{k}_{\qd_1} \ket{k}_{\p_3} \ket{i}_{\p_1} \ket{l}_{\qd_2} \ket{l}_{\p_4}\ket{j}_{\p_2}$
   
  \item[Step 8.] Perform an $H$ gate on each emitter to get  \\
    $\hookrightarrow \ket{\phi\step8} = \\ \sum_{\substack{i,j,k, \\ l,m,r}}  \Omega \ \ket{m}_{\qd_1} \ket{k}_{\p_3} \ket{i}_{\p_1} \ket{r}_{\qd_2} \ket{l}_{\p_4}\ket{j}_{\p_2}$\\
    where $\Omega \coloneqq \omega^{ij}  \omega^{ik} \omega^{jl} \omega^{il} \omega^{kl} \omega^{km} \omega^{lr}$
    
   \item[Step 9.] Perform three $CZ$ gates: One $CZ$ on $\p_4$ and $\qd_1$, the other $CZ$ on $\p_2$ and $\qd_2$, and the third $CZ$ operator on $\qd_1$ and $\qd_2$. With this we obtain\\
   $\hookrightarrow \ket{\phi\step9} =$
   \[\sum_{\substack{i,j,k, \\ l,m,r=0}}^{\dloc-1} \Omega' \ \ket{m}_{\qd_1} \ket{k}_{\p_3} \ket{i}_{\p_1} \ket{r}_{\qd_2} \ket{l}_{\p_4} \ket{j}_{\p_2},
      \]
   where $\Omega' \coloneqq \omega^{ij}  \omega^{ik} \omega^{jl} \omega^{il} \omega^{kl} \omega^{km} \omega^{lr}\omega^{lm} \omega^{jr} \omega^{mr}$
   
   \item[10.] Perform a photon-pumping operation on each emitter, followed by an $H$ gate on each emitter. Measure both emitters and perform the gates $Z_{\p_5}^{(q-1)o_1}Z_{\p_6}^{(q-1)o_2}$ on the newly generated photons $\p_5$ and $\p_6$, where $o_1$ and $o_2$ are the measurement outcomes. The final state is
   \begin{equation}
   \begin{split}
   &\hookrightarrow \ket{\phi\step{10}} =\ket{\AME(6,\dloc)_{\text{graph}}} = \\
   &\sum_{\substack{i,j,k, \\ l,m,r=0}}^{\dloc-1} \Omega' \ \ket{m}_{\p_5} \ket{k}_{\p_3} \ket{i}_{\p_1} \ket{r}_{\p_6} \ket{l}_{\p_4} \ket{j}_{\p_2},\nonumber
      \end{split}
   \end{equation}
   which is the one shown in Fig.~\ref{fig:AME6,q}(a).
 \end{itemize}

We now present a protocol for generating the $\AME(6,\dloc)$ states corresponding to the graph shown in Fig.~\ref{fig:AME6,q}(b). This protocol requires three coupled emitters with $\dloc$ ground levels each but does not need photon-emitter interaction: 

 \begin{itemize}[leftmargin=.5in]
   \item[Step 1.] Prepare three emitters in the state\\ 
   $\hookrightarrow \ket{\phi\step1} =\ket{0}_{\qd_1}\ket{0}_{\qd_2} \ket{0}_{\qd_3}$. 
   
   \item[Step 2.] Perform a Hadamard gate $H$, Eq.~\eqref{eq:H-on-qudits}, on each emitter to obtain\\ 
   $\hookrightarrow \ket{\phi\step2} =\sum_{i,j,k=0}^{\dloc -1} \ket{i}_{\qd_1} \ket{j}_{\qd_2} \ket{k}_{\qd_3}$.
   
   \item[Step 3.] Perform three $CZ$ gates on each pair of emitters (i.e., on $\qd_1$ and $\qd_2$, on $\qd_2$ and $\qd_3$, and on $\qd_1$ and $\qd_3$): \\
      $\hookrightarrow \ket{\phi\step3} =\sum_{i,j,k=0}^{\dloc -1}  \omega^{ij} \omega^{jk} \omega^{ik} \  \ket{i}_{\qd_1}\ket{j}_{\qd_2} \ket{k}_{\qd_3}$.
   
   \item[Step 4.] Perform photon-pumping operations ${\cal P}_\mathrm{pump}$, Eq.~\eqref{eq:photon-pumping}, to each emitter to create three photons:\\
   $\hookrightarrow \ket{\phi\step4} =\sum_{i,j,k=0}^{\dloc -1} $\\
   $ \omega^{ij} \omega^{jk} \omega^{ik} \  \ket{i}_{\qd_1} \ket{i}_{\p_1} \ket{j}_{\qd_2} \ket{j}_{\p_2} \ket{k}_{\qd_3} \ket{k}_{\p_3}$.
   
   \item[Step 5.] Perform an $H$ gate on each emitter: \\
      $\hookrightarrow \ket{\phi\step5} =$\\
      $\sum_{i,j,k,l,m,r=0}^{\dloc -1} \Theta  \  \ket{l}_{\qd_1} \ket{i}_{\p_1} \ket{m}_{\qd_2} \ket{j}_{\p_2} \ket{r}_{\qd_3} \ket{k}_{\p_3}$, 
      \\
      where $\Theta  \coloneqq \omega^{ij} \omega^{jk} \omega^{ik} \omega^{il} \omega^{jm} \omega^{kr}$.
      
     \item[Step 6.] Again perform a $CZ$ gate on each pair of emitters:\\
     $\hookrightarrow \ket{\phi\step6} =$
     \[ \sum_{i,j,k,l,m,r=0}^{\dloc -1} \Theta'  \  \ket{l}_{\qd_1} \ket{i}_{\p_1} \ket{m}_{\qd_2} \ket{j}_{\p_2} \ket{r}_{\qd_3} \ket{k}_{\p_3},
       \]     
        where 
$        \Theta'  \coloneqq \omega^{ij} \omega^{jk} \omega^{ik} \omega^{il} \omega^{jm} \omega^{kr} \omega^{lm} \omega^{mr} \omega^{lr}$.

\item[Step 7.] Perform a photon-pumping operation on each emitter, followed by an $H$ gate on each emitter. Measure all three emitters and perform the gates $Z_{\p_4}^{(q-1)o_1}Z_{\p_5}^{(q-1)o_2}Z_{\p_6}^{(q-1)o_3}$ on the newly generated photons $\p_4$, $\p_5$, and $\p_6$, where $o_1$, $o_2$, $o_3$ are the measurement outcomes. The final state is
\begin{equation}
       \begin{split}
   &\hookrightarrow \ket{\phi\step7} =\ket{\AME(6,\dloc)_{\text{graph}}} = \\
   &    \sum_{i,j,k,l,m,r=0}^{\dloc -1} \Theta'  \  \ket{l}_{\p_4} \ket{i}_{\p_1} \ket{m}_{\p_5} \ket{j}_{\p_2} \ket{r}_{\p_6} \ket{k}_{\p_3},\nonumber
       \end{split}
      \end{equation}   
       which is the one shown in Fig.~\ref{fig:AME6,q}(b). 
 \end{itemize}
The quantum circuits summarizing each protocol are shown in Fig.~\ref{fig:AME6,q}.

 \section{How to generate AME states of $7$ qutrits }\label{app:AME73}

Here, we discuss how to generate an $\AME$ state of $7$ qutrits. 
We use two quantum emitters with three ground levels each, and we allow photon-emitter interaction (four $CZ$ gates between photons and emitters).
The protocol is similar to the one presented above for the AME(6,$\dloc$) states depicted in Fig.~\ref{fig:AME6,q}(a) due to the similarity in graph structure.
The details are follows:
\begin{itemize}[leftmargin=.5in]
   \item[Step 1.] Prepare the two quantum emitters in the state\\
    $\hookrightarrow \ket{\phi\step1} =\ket{0}_{\qd_1}\ket{0}_{\qd_2}$ 
   
   \item[Step 2.] Perform a Hadamard gate $H$, Eq.~\eqref{eq:H-on-qudits}, on each emitter:\\ 
   $\hookrightarrow \ket{\phi\step2} =\ket{X_0}_{\qd_1} \ket{X_0}_{\qd_2} = \sum_{i,j=0}^{2} \ket{i}_{\qd_1}\ket{j}_{\qd_2}$
   
   \item[Step 3.] Perform $CZ^2$, Eq.~\eqref{eq:CZ2qutrit}, on the two emitters to get\\
    $\hookrightarrow \ket{\phi\step3} =\sum_{i,j=0}^{2} \omega^{2ij} \  \ket{i}_{\qd_1}\ket{j}_{\qd_2}$
   
   \item[Step 4.] Perform the photon-pumping operation ${\cal P}_\mathrm{pump}$, Eq.~\eqref{eq:photon-pumping}, on each emitter to create two photons:\\
   $\hookrightarrow \ket{\phi\step4} = \sum_{i,j=0}^{2} \omega^{2ij} \  \ket{i}_{\qd_1} \ket{i}_{\p_1} \ket{j}_{\qd_2} \ket{j}_{\p_2}$
   
   \item[Step 5.] Perform an $H$ gate  on each emitter: \\
   $\hookrightarrow \ket{\phi\step5} = \\\sum_{i,j,k,l=0}^{2} \omega^{2ij}  \omega^{ik} \omega^{jl}\  \ket{k}_{\qd_1} \ket{i}_{\p_1} \ket{l}_{\qd_2} \ket{j}_{\p_2}$ 
   
   \item[Step 6.] Apply two $CZ$ gates, Eq.~\eqref{eq:CZqutrit}, one on the first photon ($\p_1$) and the second emitter ($\qd_2$), and the other on the two emitters ($\qd_1$ and $\qd_2$), yielding\\
   $\hookrightarrow \ket{\phi\step6} =$\\
   $ \sum_{i,j,k,l=0}^{2} \omega^{2ij}  \omega^{ik} \omega^{jl} \omega^{il} \omega^{kl}\  \ket{k}_{\qd_1} \ket{i}_{\p_1} \ket{l}_{\qd_2} \ket{j}_{\p_2}$
    
   \item[Step 7.] Perform photon-pumping operations on each emitter to create two more photons: \\
   $\hookrightarrow \ket{\phi\step7} = $
   \\$  \sum_{\substack{i,j, \\ k,l}}\omega^{2ij}  \omega^{ik} \omega^{jl} \omega^{il} \omega^{kl}\ \ket{k}_{\qd_1} \ket{k}_{\p_3} \ket{i}_{\p_1} \ket{l}_{\qd_2} \ket{l}_{\p_4}\ket{j}_{\p_2}$
   
  \item[Step 8.] Perform an $H$ gate on each emitter: \\
  $\hookrightarrow \ket{\phi\step8} = $\\
  $\sum_{\substack{i,j,k, \\ l,m,r=0}}^2  \Xi \ \ket{m}_{\qd_1} \ket{k}_{\p_3} \ket{i}_{\p_1} \ket{r}_{\qd_2} \ket{l}_{\p_4}\ket{j}_{\p_2}$\\
   where $\Xi \coloneqq \omega^{2ij}  \omega^{ik} \omega^{jl} \omega^{il} \omega^{kl} \omega^{km} \omega^{lr}$
   
   \item[Step 9.] Perform two $CZ$ gates, one on $\p_4$ and $\qd_1$, the other on $\qd_1$ and $\qd_2$, yielding\\
   $\hookrightarrow \ket{\phi\step9} =\\\sum_{\substack{i,j,k, \\ l,m,r=0}}^{2} \Xi' \ \ket{m}_{\qd_1} \ket{k}_{\p_3} \ket{i}_{\p_1} \ket{r}_{\qd_2} \ket{l}_{\p_4} \ket{j}_{\p_2}$\\
   where $\Xi' \coloneqq \omega^{2ij}  \omega^{ik} \omega^{jl} \omega^{il} \omega^{kl} \omega^{km} \omega^{lr}\omega^{lm} \omega^{mr}$
   
   \item[Step 10.] Perform a photon-pumping operation on only $\qd_2$ to create one more photon:\\
      $\hookrightarrow \ket{\phi\step{10}} =$\\
      $\sum_{\substack{i,j,k, \\ l,m,r=0}}^{2} \Xi' \ \ket{m}_{\qd_1} \ket{k}_{\p_3} \ket{i}_{\p_1} \ket{r}_{\qd_2} \ket{r}_{\p_5} \ket{l}_{\p_4} \ket{j}_{\p_2}$
      
   \item[Step 11.] Perform an $H^\dag$ gate, Eq.~\eqref{eq:Hdag-on-qudits},  on $\qd_2$ to yield \\
           $\hookrightarrow \ket{\phi\step{11}} =$\\
           $\sum_{\substack{i,j,k,l \\ m,r,s=0}}^{2} \Xi'' \ \ket{m}_{\qd_1} \ket{k}_{\p_3} \ket{i}_{\p_1} \ket{s}_{\qd_2} \ket{r}_{\p_5} \ket{l}_{\p_4} \ket{j}_{\p_2}$\\
   where $\Xi'' \coloneqq \omega^{2ij}  \omega^{ik} \omega^{jl} \omega^{il} \omega^{kl} \omega^{km} \omega^{lr}\omega^{lm} \omega^{mr} \omega^{2sr}$
   
    \item[Step 12.] Apply two $CZ$ gates, one on the fourth photon ($\p_4$) and the second emitter ($\qd_2$), and the other on $\p_2$ and $\qd_2$:\\
    $\hookrightarrow \ket{\phi\step{12}} =$
      \[\sum_{\substack{i,j,k,l \\ m,r,s=0}}^{2} \tilde{\Xi} \ \ket{m}_{\qd_1} \ket{k}_{\p_3} \ket{i}_{\p_1} \ket{s}_{\qd_2} \ket{r}_{\p_5}  \ket{l}_{\p_4} \ket{j}_{\p_2}
      \]
         where 
\[\tilde{\Xi} \coloneqq \omega^{2ij}\omega^{ik} \omega^{jl} \omega^{il} \omega^{kl} \omega^{km} \omega^{lr} \omega^{lm} \omega^{mr} \omega^{2sr} \omega^{sl} \omega^{js}\]

\item[Step 13.] Perform a photon-pumping operation on each emitter, followed by an $H$ gate on each emitter. Measure both emitters in the $Z$ basis and perform the gates $Z_{\p_6}^{(q-1)o_1}Z_{\p_7}^{(q-1)o_2}$ on the newly generated photons, $\p_6$ and $\p_7$, where $o_1$ and $o_2$ are the measurement outcomes. The final state is\\
\begin{equation}
   \begin{split}
   &\hookrightarrow \ket{\phi\step{13}} =\ket{\AME(7,3)_{\text{graph}}} = \\
   &\sum_{\substack{i,j,k,l \\ m,r,s=0}}^{2} \tilde{\Xi} \ \ket{m}_{\p_6} \ket{k}_{\p_3} \ket{i}_{\p_1} \ket{s}_{\p_7} \ket{r}_{\p_5}  \ket{l}_{\p_4} \ket{j}_{\p_2},\nonumber
      \end{split}
   \end{equation}
   which is the state shown in Fig.~\ref{fig:AME7,3}(a).
 \end{itemize}
 
 The quantum circuit for constructing $\ket{\AME(7,3)_{\text{graph}}}$ is presented in Fig.~\ref{fig:AME7,3}(b). 
 
}

\bibliography{AME}
\end{document}